\documentclass[a4paper,12pt]{article}
 
\pdfoutput=1
\usepackage{jheppub}
\usepackage{dcolumn}
\usepackage{soul}
\usepackage[english]{babel}
\usepackage[utf8]{inputenc}
\usepackage{dcolumn}
\usepackage{soul}
\usepackage{hyperref}
\usepackage{url}
\usepackage{graphicx}
\usepackage{bm,amsmath,amssymb}
\usepackage[mathscr]{eucal}
\usepackage{makeidx}
\usepackage{subfig}
\usepackage{amsmath}
\usepackage{amssymb}
\usepackage{amsthm}
\usepackage{mathrsfs}
\usepackage{graphicx}
\usepackage{fancyhdr}
\usepackage{array}
\usepackage{simplewick}
\usepackage{latexsym}
\usepackage[all]{xy}
\usepackage{enumerate}
\usepackage{dsfont}
\usepackage{slashed}
\usepackage{float}
\usepackage[titletoc]{appendix}
\usepackage{ulem}
\usepackage{physics} 

\usepackage{verbatim}

\usepackage{tikz,tikz-3dplot}
\tdplotsetmaincoords{80}{45}

\tikzset{surface/.style={draw=black, fill=white, fill opacity=.6}}

\usepackage{tikz}
\usetikzlibrary{decorations.pathmorphing,patterns}

\newcommand{\be}{\begin{equation}}
\newcommand{\ee}{\end{equation}}
\newcommand{\bea}{\begin{eqnarray}}
\newcommand{\eea}{\end{eqnarray}}

\newcommand{\ben}{\begin{eqnarray}}
\newcommand{\een}{\end{eqnarray}}

\title{\boldmath\LARGE{{Black Hole Entropy and Complexity Growth in Horndeski Gravity within the AdS/BCFT Framework}}}

\author[a]{Fabiano F. Santos,}
\author[b]{Behnam Pourhassan,}
\author[c]{Emmanuel N. Saridakis}
\affiliation[a,b]{School of Physics, Damghan University, Damghan, 3671641167, Iran.}
\affiliation[a]{Centro de Ciências Exatas, Naturais e Tecnológicas, UEMASUL, 65901-480, Imperatriz, MA, Brazil.\\Departamento de Física, Universidade Federal do Maranhão, São Luís, 65080-805, Brazil.}
\affiliation[c]{National Observatory of Athens, Lofos Nymfon 11852, Greece.\\
CAS Key Laboratory for Research in Galaxies and Cosmology, School of Astronomy and Space Science, University of Science and Technology of China, Hefei 230026, China.\\
Departamento de Matem\'{a}ticas, Universidad Cat\'{o}lica del Norte, Avda. Angamos 0610, Casilla 1280, Antofagasta, Chile.}
\emailAdd{fabiano.ffs23@gmail.com}
\emailAdd{b.pourhassan@du.ac.ir}
\emailAdd{msaridak@noa.gr}

\abstract{This work investigates the connection between quantum complexity and gravitational dynamics within the framework of Horndeski gravity, extending the AdS/BCFT correspondence to include scalar-tensor interactions. By refining the “complexity = action” conjecture, we analyze how Horndeski gravity modifies the Wheeler-DeWitt patch and the causal structure relevant for holographic complexity. Our analysis shows that the linear growth of complexity, proportional to the product of black hole entropy and temperature, is recovered for the class of black hole configurations studied here, including rotating and charged solutions. We also study the effect of shock waves on complexity growth and find the appearance of the switchback effect. These conclusions hold in the regime in which the relevant effective characteristic cone of Horndeski gravity is compatible with the null structure used to define the Wheeler-DeWitt patch, so that the causal construction effectively matches that of the background metric. Within this domain of validity, our results provide evidence for the complexity = action conjecture in Horndeski gravity and illustrate how modified gravitational couplings affect holographic complexity in the AdS/BCFT framework.}

\begin{document}
	\maketitle
	\newcommand{\limit}[3]
	{\ensuremath{\lim_{#1 \rightarrow #2} #3}}

\section{Introduction}

The interior of a black hole is a famous example of emergent spacetime \cite{complexityshocks,entrenholo,vidal_tns_geo,Hartman:2013qma,Susskind:2014moa,Santos:2024cvx}, characterized by its linear growth over exponentially long timescales after the black hole's formation \cite{Hartman:2013qma,Susskind:2014moa}. Within the framework of holography, a known conjecture posits that this growth is dual to the evolution of quantum complexity in the boundary theory \cite{Susskind:2014rva,Brown:2015bva,Lloyd:2000cry,Brown:2015lvg,Susskind:2018fmx,Brown:2018bms,Brown:2017jil,Brown:2019whu,Brown:2022rwi}. Although this duality remains a conjecture, it has undergone extensive testing and has shown remarkable consistency with theoretical predictions.

In the context of the Anti-de Sitter/Conformal Field Theory (AdS/CFT) correspondence \cite{Maldacena:1997re,Witten:1998qj,Karch:2025fky,Geng:2023iqd,Geng:2020qvw,Geng:2020fxl,Geng:2021wcq,Geng:2021iyq,Geng:2021mic,Geng:2022dua,Geng:2023qwm,Geng:2022slq,Geng:2022tfc,Geng:2024xpj,Geng:2025efs,Bao:2025plr,Geng:2025yys,Harper:2024aku,Kawamoto:2023wzj,Aharony:2008wz,Bousso:2024ysg,Akal:2020ujg,Cardy:2008jc,Calabrese:2004eu,1,2,3,4,5,6}, this conjecture has been formalized through the "$complexity=volume$" ${\cal C}{\cal V}$ proposal, which asserts that the volume of a maximal space-like slice extending into the black hole interior is proportional to the computational complexity of the boundary Conformal Field Theory (CFT) state \cite{complexityshocks}. This aligns with the broader connection between tensor networks and geometry, where the geometry of the bulk spacetime is associated with the minimal tensor network required to prepare the boundary state \cite{entrenholo,vidal_tns_geo,Hartman:2013qma,Santos:2024cvx}.

Extending this framework to the Anti-de Sitter/Boundary Conformal Field Theory (AdS/BCFT) correspondence \cite{Santos:2024cwf,Takayanagi:2011zk,Santos:2024cvx,Braccia:2019xxi,Santos:2023eqp,Santos:2025wbl,Hao:2025ocu,Fujiki:2025yyf,Kanda:2023zse,Izumi:2022opi,Kawamoto:2022etl,Suzuki:2022xwv,Miyaji:2021ktr,Takayanagi:2020njm,Miyaji:2016mxg,Miyagawa:2015sql,Nozaki:2013wia,Nozaki:2012qd}, where the bulk spacetime is dual to a boundary CFT with tension, is an interesting physical scenario. In particular, the Horndeski gravity \cite{Horndeski:2024sjk}, a modified gravity theory beyond Einstein gravity with second-derivative scalar-tensor interactions  \cite{CANTATA:2021asi}, modifies the bulk dynamics and boundary conditions \cite{Santos:2024cvx,Santos:2023eqp}. These modifications have profound implications for the geometry of black hole interiors and the dual description of quantum complexity \cite{Santos:2024zoh,Santos:2020xox}. For instance, Horndeski gravity can alter the causal structure of the bulk spacetime, potentially leading to new insights into the growth of complexity and its holographic interpretation.

In the context of the two-sided AdS black hole \cite{Brown:2015bva,Brown:2015lvg,Luciano:2023bai}, the relationship between quantum complexity and geometry has been conjectured to take the form:
\begin{eqnarray}
\text{Complexity} \sim \frac{{\cal V}}{G_{N}\,l_{AdS}}, \nonumber
\end{eqnarray}
where \( {\cal V} \) represents the volume of the Einstein-Rosen bridge (ERB) \cite{Susskind:2014yaa}, \( l_{AdS} \) is the radius of the AdS curvature and \( G_{N} \) is Newton's constant. However, a new perspective on the complexity and geometry is given by:
\begin{eqnarray}
\text{Complexity} \sim \frac{{\cal C}\,{\cal W}}{G_N\,l^2_{AdS}}.\nonumber
\end{eqnarray}
Here, \( {\cal C}{\cal W}=l_{Ads} {\cal V} \) is the spacetime volume of the ERB. This reformulation highlights the role of the cosmological constant \( \Lambda \sim -1/\l^2_{AdS}\), in connecting quantum information and gravitational dynamics \cite{Brown:2015bva,Brown:2015lvg}. Inspired by this observation, a new conjecture, known as the Complexity-Action (${\cal C}{\cal A}$) conjecture, has been proposed:

\begin{eqnarray}
\text{${\cal C}{\cal A}$-conjecture:}\,\,{\cal C}(t)=\frac{d{\cal C}}{dt} = \frac{2\dot{{\cal A}}(t)}{\pi \hbar}\,\,;\,\,\dot{{\cal A}}(t)=\frac{d{\cal\,A}ction}{dt}.\nonumber   
\end{eqnarray}
This conjecture suggests that the quantum complexity of a boundary state is proportional to the gravitational action evaluated in the bulk spacetime \cite{Brown:2015bva}. 

In AdS/CFT this bulk region is usually taken to be the Wheeler–DeWitt (WdW) patch: the domain of dependence of a bulk Cauchy slice anchored at boundary times $t_L$ and $t_R$ \cite{Brown:2015bva,Brown:2015lvg}. In Einstein gravity the boundary of this domain is null, so the WdW patch is naturally constructed using null rays. However, in Horndeski theories the causal structure is generically richer and more subtle. Different graviton polarizations and scalar degrees of freedom can propagate with different effective speeds, and characteristic hypersurfaces need not coincide with the null cones of the physical metric. In particular, characteristic hypersurfaces can become spacelike, corresponding to superluminal modes, while the initial–value problem can nevertheless remain well posed \cite{Kovacs:2020pns,Papallo:2017qvl,Bettoni:2016mij,Bruyere:2026gnt}. Consequently, in general Horndeski models the boundary of the domain of dependence – and hence the WdW patch – is not a priori determined by the null rays of the Einstein frame metric. A careful treatment of holographic complexity in Horndeski gravity therefore requires specifying which characteristic cone sets the causal domain that defines the WdW patch. Under suitable symmetry and parameter restrictions one can define an effective metric with respect to which the relevant modes propagate at the maximal speed, and in particular regimes this effective characteristic cone can coincide with, or lie close to, the null cone of the physical metric \cite{Papallo:2017qvl,Bettoni:2016mij}. In this work we will focus on such regimes, in which the fastest propagating modes are effectively governed by a causal structure that can be described by null hypersurfaces of an appropriately chosen metric. We will comment Sec. \ref{v0} and Sec. \ref{TEST} on how this assumption enters our constructions and how it relates to the broader literature on Horndeski causal cones and well‑posedness.

\begin{figure}[ht]
\centering
\includegraphics[scale=0.65]{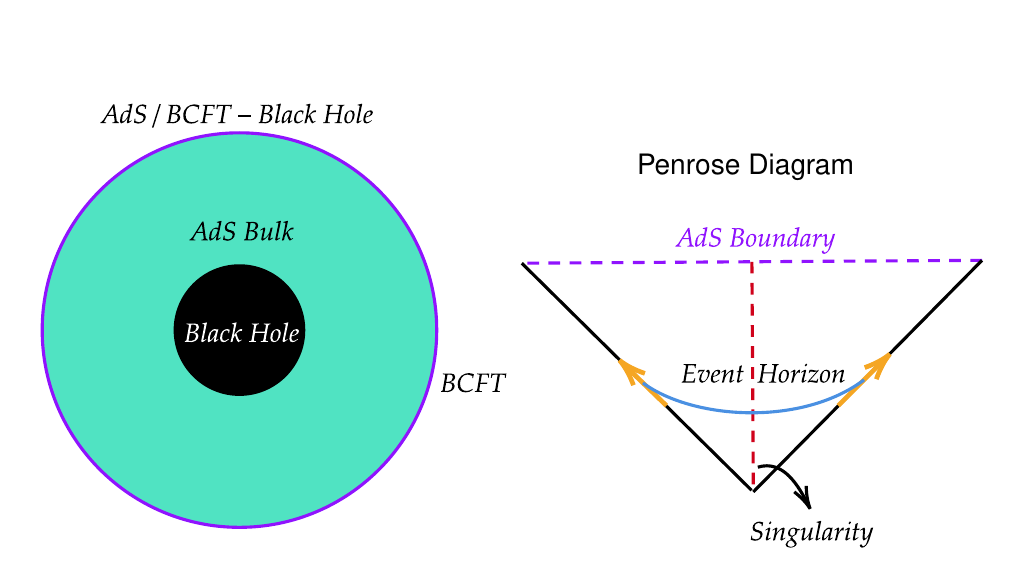}
\caption{{\sl Left side}, AdS bulk is represented as a hyperbolic space, with the black hole located at its center. The boundary of the AdS space hosts the boundary conformal field theory (BCFT), which encodes the dual description of the bulk gravitational dynamics. {\sl Right side}, the Penrose diagram illustrates the causal structure of the black hole, including the event horizon, singularity, and AdS boundary.}
\label{fig:stream}
\end{figure}

To further explore this connection, we consider the AdS/BCFT correspondence \cite{Santos:2024cwf,Takayanagi:2011zk,Santos:2024cvx,Braccia:2019xxi}, which is an extension of the AdS/CFT including a boundary Fig. \ref{fig:stream}. In the framework of Horndeski gravity \cite{Santos:2024cvx,Horndeski:2024sjk,Santos:2024zoh,Santos:2020xox,Santos:2021orr,Santos:2023flb,Feng:2015oea,Hajian:2020dcq,Feng:2015wvb,Leon:2012mt,Bahamonde:2021dqn}, which extends Einstein's theory \cite{Braccia:2019xxi} by including the derivatives of scalar fields that can couple directly to the Einstein tensor ($\gamma\,G^{\mu\nu}\nabla_{\mu}\phi\nabla_{\nu}\phi$), leading to a broader class of scalar-tensor theories \cite{Iyer:1994ys,Barnich:2007bf,Khodabakhshi:2020fhb,Caceres:2017lbr,Dong:2013qoa,Minamitsuji:2013vra,Callan:2012ip,Hubeny:2013gta,Tsilioukas:2024seh,Tsilioukas:2025dmy}. The dynamics of the brane and its embedding in the bulk spacetime provide a rich scenario for studying the relationship between complexity, geometry, and quantum information \cite{Santos:2024cvx}. Specifically, the interaction between the extrinsic curvature of the brane and the bulk action may offer new insight into conjecture ${\cal C}{\cal A}$ and its implications for holography \cite{Brown:2015bva,Brown:2015lvg}.

\begin{figure}[ht]
\centering
\includegraphics[scale=0.75]{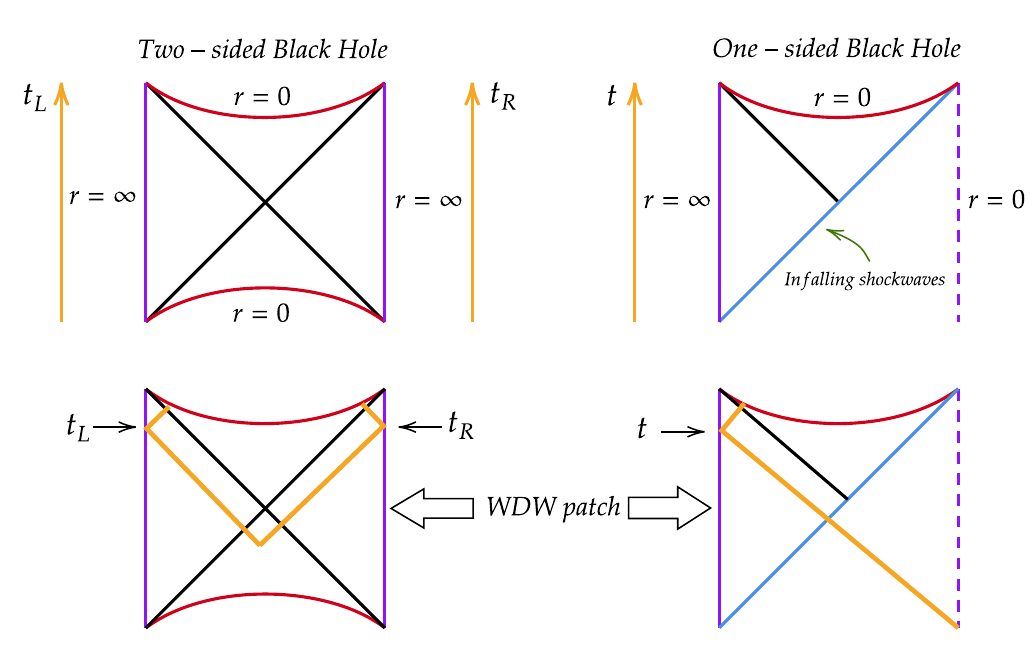}
\caption{{\sl In the right side}, eternal two-sided Anti-de Sitter (AdS) black hole and its one-sided counterpart formed through shockwave collapse represent distinct holographic frameworks with profound implications for Boundary Conformal Field Theories (BCFTs). The two-sided configuration corresponds to an entangled thermofield double state across dual CFTs at opposing boundaries. {\sl In the left side}, the one-sided variant maps to a single BCFT system.}
\label{fig:stream1}
\end{figure}

In the AdS/CFT framework, the Wheeler-DeWitt (WdW) patch plays a role in understanding the duality between bulk spacetime and boundary states. The WdW patch, defined as the union of all spatial slices anchored at a specific boundary time $t$ (or a pair of times ($t_L,t_R$) in the two-sided) Fig. \ref{fig:stream1}, is conjectured to encode the complexity of the boundary state through its action \cite{Brown:2015bva,Brown:2015lvg}. 

This concept can be extended to the AdS/BCFT correspondence, where the bulk spacetime includes a boundary that intersects the conformal boundary of the AdS space. In this context, the WdW patch provides a geometric representation of the complexity of states in boundary conformal field theories (BCFT) \cite{Santos:2024zoh}. Analyzing the framework of Horndeski gravity, the WdW patch offers new insights into the relationship between bulk dynamics and boundary complexity. Recent studies \cite{Santos:2024zoh,Santos:2020xox}, have explored how Horndeski gravity modifies the structure of the WdW patch and its implications for holographic complexity, suggesting novel connections between scalar-tensor theories and the AdS/BCFT correspondence.

Our refined conjecture of ``$complexity=action$'' ${\cal C}{\cal A}$ extends beyond these established paradigms because it includes boundary degrees of freedom through the AdS/BCFT correspondence \cite{Horndeski:2024sjk,Santos:2024zoh,Santos:2020xox}. This extension allows us to relate the computational complexity of the boundary theory state to the gravitational action evaluated on the Wheeler-DeWitt patch \cite{Brown:2015bva}, while accounting for additional scalar-tensor couplings that modify both bulk dynamics and boundary conditions. The non-minimal coupling terms in Horndeski gravity introduce novel contributions to holographic complexity \cite{Lloyd:2000cry}.

In order to show that the complexity is satisfied as the product between temperature and entropy of the black hole in the present scenario of this work, we propose the derivation of black hole entropy in Horndeski gravity, which has contributions from both the bulk and boundary terms. The Wald entropy formula, the Wald formalism, and the holographic renormalization scheme are used to calculate the entropy, ensuring consistency with the first law of thermodynamics \cite{Iyer:1994ys,Barnich:2007bf,Khodabakhshi:2020fhb,Caceres:2017lbr,Dong:2013qoa,Minamitsuji:2013vra,Callan:2012ip,Hubeny:2013gta}. These methods reveal that the entropy includes corrections proportional to the squared norm of the scalar fields, highlighting the role of scalar-tensor couplings in the thermodynamic properties of black holes. The study of black hole thermodynamics and quantum complexity has become a cornerstone in gravitational dynamics and quantum information theory \cite{Santos:2024cvx,Horndeski:2024sjk,Santos:2024zoh,Santos:2020xox,Santos:2021orr,Santos:2023flb,Feng:2015oea,Hajian:2020dcq,Feng:2015wvb}. Within the framework of the AdS/CFT correspondence, the "$complexity = action$" ${\cal C}{\cal A}$ conjecture has emerged as a powerful tool to connect the computational complexity of boundary conformal field theories (CFTs) with the gravitational action evaluated on the Wheeler-DeWitt (WdW) patch. This duality provides a geometric interpretation of quantum complexity, offering insights into the growth of complexity in black hole systems and its saturation at late times \cite{Susskind:2014rva,Brown:2015bva,Lloyd:2000cry,Brown:2015lvg,Susskind:2018fmx,Brown:2018bms,Brown:2017jil,Brown:2019whu,Brown:2022rwi}.

The relation between entropy, temperature, and complexity growth is further enriched by including angular momentum, electric charge, and scalar-tensor interactions \cite{Feng:2015oea,Hajian:2020dcq,Feng:2015wvb}. For rotating and charged black holes, these factors introduce additional corrections to the WdW patch, which changes the rate of complexity growth \cite{Hajian:2020dcq}. The study of shock waves and their impact on complexity growth provides a stringent test of conjecture ${\cal C}{\cal A}$, demonstrating its robustness in the presence of Horndeski modifications \cite{Brown:2015lvg}.

This work also explores the connection between entanglement entropy, phase transitions, and complexity growth in the context of Horndeski gravity. The functional entanglement entropy \cite{Caceres:2017lbr}, derived for both planar and spherical BTZ black holes, reflects the intricate relationship between bulk geometry and boundary degrees of freedom. The phase transition between connected and disconnected minimal surfaces provides a geometric manifestation of the transition from linear complexity growth to saturation, offering a deeper understanding of the thermodynamic and quantum information properties of black hole systems.\\

The new contributions of this paper are instead:\\

$\bullet$ We embed Horndeski black holes into the AdS/BCFT setup and analyze black hole entropy and complexity growth in this framework, including explicit boundary contributions and their dependence on the Horndeski couplings.\\

$\bullet$ We provide a unified derivation showing that, for the class of Horndeski models and parameter regimes we consider, the CA conjecture yields a linear growth rate of complexity proportional to \(T S_{BH}\) across several examples (planar AdS black holes, rotating BTZ black holes, and charged AdS black holes with Horndeski corrections), thus highlighting a form of universality of the CA duality in modified gravity.\\

$\bullet$ In the BCFT context, we study the entanglement entropy and boundary entropy, including the dependence on the Horndeski coupling \(\gamma\), and we relate the phase transition between connected and disconnected minimal surfaces in the spherical BTZ case to the transition from linear complexity growth to saturation.\\

$\bullet$ We analyze the impact of shock waves and the switchback effect on complexity growth in Horndeski–AdS/BCFT geometries, discussing how Horndeski modifications of the causal structure enter the CA calculation and under which assumptions the standard shock‑wave intuition continues to hold.\\

This work is summarized as follows. In Sec. \ref{v0}, we present our physical system of the AdS/BCFT within Horndeski gravity. In Sec. \ref{WEHT}, we present three ways to compute the entropy in the Horndeski gravity framework: the Wald entropy formula (\ref{W1}), the Wald formalism (\ref{W2}), and the holographic renormalization scheme (\ref{W3}). In Sec. \ref{NU}, we present the phase transition between connected and disconnected minimal surfaces in spherical BTZ black holes, representing the shift from linear complexity growth (\(\theta_0 < \theta_c\)) to complexity saturation as \(\theta_0\) approaches \(\theta_c\). In Sec. \ref{BHSL}, we present the product between temperature and entropy for three different kinds of black holes: AdS black hole (\ref{ADSBH}), the BTZ black hole with angular momentum (\ref{BH-BTZ}), and the charge AdS black hole (\ref{BH-EHW}). For all these cases, the conjecture "$complexity = action$" ${\cal C}{\cal A}$ is satisfied. In Sec. \ref{TEST}, we show that the description of the shock waves is satisfied and that their impact on the Horndeski gravity on complexity growth provides a stringent test of the ${\cal C}{\cal A}$ conjecture \cite{Brown:2015lvg}. Finally, in Section \ref{CONC}, we provide our conclusions and discuss the results of this work.

\section{Horndeski Gravity within the AdS/BCFT Framework}\label{v0}
\begin{figure}[ht]
\centering
\includegraphics[scale=0.75]{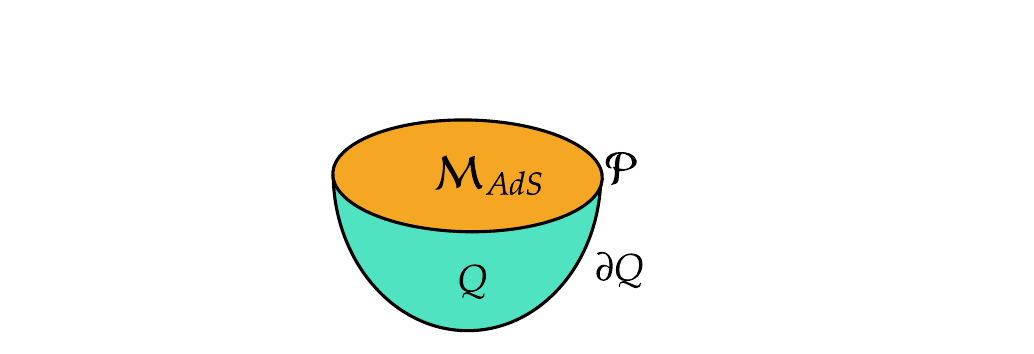}
\caption{AdS/CFT correspondence in the presence of boundary hypersurface.}
\label{fig:stream2}
\end{figure}

Now we present the Horndeski Gravity within the AdS/BCFT Framework \cite{Horndeski:2024sjk,Santos:2024zoh,Santos:2020xox}. To construct this system, we will consider those whose bulk lagrangian ${\cal L}^{{\cal\,Q}}_{H}$ has a contribution of the boundary lagrangian ${\cal L}^{\partial{\cal\,Q}}_{bdry}$. So, the total action can be written as
\begin{eqnarray}
&&\mathcal{A}={\cal\,A}^{\mathcal{Q}}_{\rm H}+{\cal\,A}^{\partial\mathcal{Q}}_{bdry}=\frac{1}{\kappa}\int_{\mathcal{Q}}{\sqrt{|g|}{\cal L}^{\mathcal{Q}}_{\rm H}}+\int_{\mathcal{Q}}{\sqrt{|g|}{\cal L}^{\mathcal{Q}}_{\rm {\cal M}}}+\frac{2}{\kappa}\int_{\partial\mathcal{Q}}{\sqrt{|h|}\mathcal{L}^{\partial\mathcal{Q}}_{bdry}}\cr
&&\,\,\,\,\,\,\,\,\,\,\,\,\,\,\,\,\,\,\,\,\,\,\,\,\,\,\,\,\,\,\,\,\,\,\,\,\,\,\,\,\,\,\,\,\,\,+\int_{\partial\mathcal{Q}}{\sqrt{|h|}\mathcal{L}^{\partial\mathcal{Q}}_{mat}}+\frac{2}{\kappa}\int_{ct}{\sqrt{|h|}\mathcal{L}^{\mathcal{P}}_{ct}},\label{1}
\end{eqnarray}
with 
{\begin{eqnarray}
&&{\cal L}^{\mathcal{Q}}_{\rm H}= (R-2\Lambda)\label{L1} -\frac{1}{2}(\alpha g_{\mu\nu}-\gamma\,  G_{\mu\nu})\nabla^{\mu}\phi\nabla^{\nu}\phi,\\
&&\mathcal{L}^{\partial\mathcal{Q}}_{bdry}=(K-\Sigma)-\frac{\gamma}{4}\nabla_{\mu}[\phi{\cal G}^{\mu\nu}],\label{L5}\\
&&{\cal G}^{\mu\nu}=(\nabla_{\nu}\phi n^{\mu}n^{\nu}-\nabla_{\nu}\phi)K-\nabla_{\nu}\phi K^{\mu\nu},\\
&&\mathcal{L}^{\mathcal{P}}_{ct}=c_{0}+c_{1}R+c_{2}R^{ij}R_{ij}+c_{3}R^{2}+b_{1}(\partial_{i}\phi\partial^{i}\phi)^{2}+\cdots,\label{L6}
\end{eqnarray}}
In the action (\ref{1}), we have contributions from the scalar curvature \( R \), the Einstein tensor \( G_{\mu \nu} \), and the cosmological constant \( \Lambda \), alongside a scalar field \( \phi = \phi(r) \) through the coupling constants \( \alpha \) and \( \gamma \). The gravitational coupling is defined as \( \kappa =16 \pi G_N\). The boundary Lagrangian, \( \mathcal{L}^{{\cal\partial\,Q}}_{\text{bdry}} \), includes the Gibbons-Hawking terms, which depend on \( \gamma \) and are essential for the consistency of Horndeski gravity. In addition, the extrinsic curvature is given by \( K_{\mu \nu} = h^{\phantom{\mu}\beta}_{\mu} \nabla_{\beta} n_{\nu} \), its trace \( K = h^{\mu \nu} K_{\mu \nu} \), the induced metric \( h_{\mu \nu} \), and the normal unit of outward pointing \( n^{\mu} \) to the boundary hypersurface \( \partial\mathcal{Q} \) Fig. \ref{fig:stream2}. Additionally, \( \Sigma \) represents the boundary tension in \( \partial\mathcal{Q} \), while \( \mathcal{L}^{\partial\mathcal{Q}}_{\text{mat}} \) denotes the Lagrangian matter in \( \partial\mathcal{Q} \). The counterterms \( \mathcal{L}^{\mathcal{P}}_{\text{ct}} \), which are introduced to regularize divergences.

This formulation is particularly relevant in the context of the AdS/BCFT correspondence \cite{Santos:2024cwf,Takayanagi:2011zk}, where the correspondence between bulk and boundary dynamics is established. The inclusion of Horndeski gravity enables a richer structure in the boundary conditions, which can be used to explore novel features of the correspondence. For example, the boundary terms \( \mathcal{L}^{{\cal\partial\,Q}}_{\text{bdry}}\) and \( \mathcal{L}_{\text{mat}} \) can be interpreted as encoding the holographic dual of conformal field theory (CFT) living on the boundary. Recent studies, such as those in \cite{Litvinova:2015jba,Frey:2013bha}, have highlighted the role of boundary counterterms and extrinsic curvature in the AdS/BCFT framework, particularly when extended to modified gravity theories, including Horndeski gravity.

Following the procedures outlined in \cite{Santos:2024cwf,Takayanagi:2011zk,Santos:2024cvx}, and imposing Neumann boundary conditions on the boundary of the bulk spacetime, the equations of motion for the boundary are obtained varying the boundary action (\( \delta{\cal\,A}^{\partial\mathcal{Q}}_{bdry}=0 \)):
\begin{eqnarray}
K_{\alpha\beta}-h_{\alpha\beta}(K-\Sigma)+{\cal H}_{\alpha\beta}=\kappa {\cal S}^{\partial\mathcal{Q}}_{\alpha\beta}\,,\label{L7}
\end{eqnarray}
with
\begin{eqnarray}
{\cal S}^{\partial\mathcal{Q}}_{\alpha\beta}=-\frac{2}{\sqrt{-h}}\frac{\delta S^{\partial\mathcal{Q}}_{mat}}{\delta h^{\alpha\beta}}\,\,;\,\,{\cal H}_{\alpha\beta}=-\frac{2\gamma}{\sqrt{-h}}\frac{\delta(\nabla_{\mu}[\phi{\cal G}^{\mu\nu}])}{\delta h^{\alpha\beta}}.\label{L9} 
\end{eqnarray}
Taking into account the stress-energy tensor of matter in $\partial\mathcal{Q}$ as a constant (this is ${\cal S}^{\partial\mathcal{Q}}_{\alpha\beta}=0$), the equation (\ref{L7}) can be rewritten for
\begin{eqnarray}
K_{\alpha\beta}-h_{\alpha\beta}(K-\Sigma)+{\cal H}_{\alpha\beta}=0\,.\label{L10}
\end{eqnarray}
Now, for the gravity side, we vary the bulk action (\(\delta {\cal\,A}^{\mathcal{Q}}_{\rm H} \)=0):
\begin{eqnarray}
{\cal E}^{\mathcal{Q}}_{\mu\nu}=-\frac{2}{\sqrt{-g}}\frac{\delta {\cal\,A}^{\mathcal{Q}}_{H}}{\delta g^{\mu\nu}}\,,\nonumber\\
\end{eqnarray}
where the bulk equations of motion are given by:
\begin{eqnarray}
{\cal E}^{\mathcal{Q}}_{\mu\nu}&=&G_{\mu\nu}+\Lambda g_{\mu\nu}-\frac{\alpha}{2}\left(\nabla_{\mu}\phi\nabla_{\nu}\phi-\frac{1}{2}g_{\mu\nu}\nabla_{\lambda}\phi\nabla^{\lambda}\phi\right)\label{11}\nonumber\\
                  &+&\frac{\gamma}{2}\left(\frac{1}{2}\nabla_{\mu}\phi\nabla_{\nu}\phi R-2\nabla_{\lambda}\phi\nabla_{(\mu}\phi R^{\lambda}_{\nu)}-\nabla^{\lambda}\phi\nabla^{\rho}\phi R_{\mu\lambda\nu\rho}\right)\nonumber\\
									&+&\frac{\gamma}{2}\left(-(\nabla_{\mu}\nabla^{\lambda}\phi)(\nabla_{\nu}\nabla_{\lambda}\phi)+(\nabla_{\mu}\nabla_{\nu}\phi)\Box\phi+\frac{1}{2}G_{\mu\nu}(\nabla\phi)^{2}\right)\nonumber\\
									&-&\frac{\gamma g_{\mu\nu}}{2}\left(-\frac{1}{2}(\nabla^{\lambda}\nabla^{\rho}\phi)(\nabla_{\lambda}\nabla_{\rho}\phi)+\frac{1}{2}(\Box\phi)^{2}-(\nabla_{\lambda}\phi\nabla_{\rho}\phi)R^{\lambda\rho}\right),
\end{eqnarray}
The variation for the scalar field including the bulk action \( {\cal\,A}^{\mathcal{Q}}_{\rm H} \), and the boundary action \( {\cal\,A}^{\partial\mathcal{Q}}_{bdry} \) with respect to the dynamical fields yields the following: 
\begin{eqnarray}
\quad {\cal E}^{\mathcal{Q}}_{\phi}=-\frac{2}{\sqrt{-g}}\frac{\delta {\cal\,A}^{\mathcal{Q}}_{H}}{\delta\phi} \,,\quad {\cal F}^{\partial\mathcal{Q}}_{\phi}=-\frac{2}{\sqrt{-h}}\frac{\delta{\cal\,A}^{\partial\mathcal{Q}}_{bdry}}{\delta\phi} \,,\nonumber\\
\end{eqnarray}
and the equations of motion for the scalar field become:
\begin{eqnarray}
&&{\cal E}^{\mathcal{Q}}_{\phi}=\nabla_{\mu}{\cal J}^{\mu}\,\,;\,\,{\cal J}^{\mu\nu}=\left(\alpha g^{\mu\nu}-\gamma G^{\mu\nu}\right)\nabla_{\nu}\phi\,,\label{L11}\\
&&{\cal F}^{\partial\mathcal{Q}}_{\phi}=-\frac{\gamma}{4}\nabla_{\mu}{\cal G}^{\mu\nu}\,\,;\,\,{\cal G}^{\mu\nu}=(\nabla_{\nu}\phi n^{\mu}n^{\nu}-\nabla_{\nu}\phi)K-\nabla_{\nu}\phi K^{\mu\nu},\label{L12}
\end{eqnarray}
where the additional terms arise from the Horndeski scalar-tensor coupling in the boundary $\partial\mathcal{Q}$. The boundary equations satisfy the condition: ${\cal E}^{\mathcal{Q}}_{\phi}={\cal F}^{\partial\mathcal{Q}}_{\phi}$, ensuring consistency between the bulk and boundary dynamics, and have a pivotal role in determining the dual conformal field theory. 

The diagram Fig. \ref{fig:stream2} illustrates the relationship between the AdS bulk, the BCFT boundary, and the role of Horndeski gravity through the $\gamma$-parameter. The bulk spacetime is governed by the Einstein-Horndeski action (${\cal L}^{\mathcal{Q}}_{\rm H}$), while the boundary conditions ($\mathcal{L}^{\partial\mathcal{Q}}_{bdry}$) encode the dual field theory's dynamics \cite{Santos:2024cvx}. The "$complexity =action$" conjecture connects the bulk hypersurface to the computational complexity of the boundary theory, providing a deeper understanding of holographic duality in the presence of scalar-tensor interactions.

In the AdS/BCFT framework, the complexity of the boundary conformal field theory can be related to the volume of a codimension-one hypersurface in the bulk spacetime, as proposed in the "$complexity = action$" conjecture \cite{Brown:2015bva,Brown:2015lvg}. In Horndeski gravity, the scalar-tensor coupling modifies the bulk geometry, which in turn affects the volume of the hypersurface anchored to the boundary. The Neumann boundary conditions ensure that the hypersurface geometry is consistent with the modified Einstein equations, leading to a holographic complexity that reflects the role of the scalar field.

\subsection{Variational principle, boundary terms, and the effective characteristic cone}

Now provide an explicit derivation of the conditions under which the Wheeler–DeWitt (WdW) patch \cite{Kovacs:2020pns,Papallo:2017qvl,Bettoni:2016mij} can be consistently represented using null hypersurfaces of an appropriate effective metric in Horndeski gravity. The bulk Horndeski Lagrangian (\ref{L1}) can be written as a sum of two pieces,

\begin{eqnarray}
\mathcal{L}_{H}^{Q}=(R-2\Lambda)-\frac{1}{2}\big(\alpha\,g^{\mu\nu}-\gamma\,G^{\mu\nu}\big)\nabla_\mu\phi\nabla_\nu\phi.    
\end{eqnarray}
It is convenient to define the symmetric tensor
\begin{eqnarray}
Z^{\mu\nu}\;\equiv\;\alpha\,g^{\mu\nu}-\gamma\,G^{\mu\nu},
\qquad\Rightarrow\qquad
\mathcal{L}_{\phi}=-\frac12 Z^{\mu\nu}\nabla_\mu\phi\nabla_\nu\phi.    
\end{eqnarray}
Varying the scalar sector with respect to \(\phi\) gives, using \(\delta(\nabla_\mu\phi)=\nabla_\mu(\delta\phi)\),
\begin{eqnarray}
&&\delta A_{\phi}^{\rm bulk}
=-\frac{1}{4\kappa}\int_Q d^dx\,\sqrt{|g|}\,Z^{\mu\nu}\,2\nabla_\mu\phi\,\nabla_\nu(\delta\phi)\cr
&&=-\frac{1}{2\kappa}\int_Q d^dx\,\sqrt{|g|}\,Z^{\mu\nu}\nabla_\mu\phi\,\nabla_\nu(\delta\phi).    
\end{eqnarray}
Integrating by parts,
\begin{eqnarray}
&&\delta A_{\phi}^{\rm bulk}
=\frac{1}{2\kappa}\int_Q d^dx\,\sqrt{|g|}\,\nabla_\mu\!\Big(Z^{\mu\nu}\nabla_\nu\phi\Big)\,\delta\phi\cr
&&\;-\;\frac{1}{2\kappa}\int_{\partial Q} d^{d-1}x\,\sqrt{|h|}\,n_\mu\,Z^{\mu\nu}\nabla_\nu\phi\,\delta\phi.    
\end{eqnarray}
This identifies the conserved current
\begin{eqnarray}
J^\mu \;\equiv\; Z^{\mu\nu}\nabla_\nu\phi
=(\alpha g^{\mu\nu}-\gamma G^{\mu\nu})\nabla_\nu\phi,    
\end{eqnarray}
and yields the bulk scalar equation of motion
\begin{eqnarray}
E^\phi=0 \qquad\Longleftrightarrow\qquad \nabla_\mu J^\mu =0,    
\end{eqnarray}
The second term shows explicitly that the scalar variation produces a boundary flux proportional to \(n_\mu J^\mu\), so a well-posed variational principle requires specifying boundary conditions (Dirichlet, Neumann, or mixed) for \(\phi\) on \(\partial Q\), or equivalently, adding boundary terms whose variation modifies/cancels this flux.

The Einstein–Hilbert term \(\int\sqrt{|g|}R\) produces the familiar boundary term containing normal derivatives of \(\delta g_{\mu\nu}\), which is canceled by the Gibbons–Hawking term \(\int_{\partial Q}\sqrt{|h|}K\). In the present theory, the Horndeski coupling \(\propto G^{\mu\nu}\nabla_\mu\phi\nabla_\nu\phi\) also contains second derivatives of the metric (through \(G^{\mu\nu}\)), and therefore its variation generates additional boundary contributions involving \(\delta K_{\mu\nu}\) (equivalently \(n\cdot\nabla\,\delta g\)).

To see this explicitly at the level needed here, note that the problematic piece arises from the variation of the Einstein tensor,
\begin{eqnarray}
\delta\!\left(\sqrt{|g|}\,G^{\mu\nu}\nabla_\mu\phi\nabla_\nu\phi\right)
\supset
\sqrt{|g|}\,\delta G^{\mu\nu}\,\nabla_\mu\phi\nabla_\nu\phi,    
\end{eqnarray}
and \(\delta G^{\mu\nu}\) contains terms of the schematic form \(\nabla\nabla(\delta g)\). In Gaussian normal coordinates adapted to \(\partial Q\), these second derivatives decompose into tangential derivatives and a normal derivative piece proportional to \(\delta K_{\mu\nu}\) via
\begin{eqnarray}
 \delta K_{\mu\nu} \;=\; \frac12\,h_\mu{}^{\rho}h_\nu{}^{\sigma}\,n^\lambda\nabla_\lambda(\delta g_{\rho\sigma})+\cdots,   
\end{eqnarray}
so \(\delta A_{H}^{Q}\) contains boundary terms of the form
\begin{eqnarray}
&&\delta A_{H}^{Q}\big|_{\partial Q}\supset
\frac{\gamma}{2\kappa}\int_{\partial Q}\! d^{d-1}x\,\sqrt{|h|}\,
\Big[\cdots\Big]^{\mu\nu}\,\delta K_{\mu\nu}\cr
&&\;+\;(\text{terms}\propto \delta h_{\mu\nu}) \;+\;(\text{terms}\propto \delta\phi).    
\end{eqnarray}
Therefore, just as \(K\) is required for Einstein–Hilbert, additional \(\gamma\)-dependent boundary contributions are required for a well-defined variational principle in the Horndeski theory.

A convenient choice (equivalent to the form used later in the Euclidean action, up to total derivatives on \(\partial Q\) and possible corner terms at \(P=\partial(\partial Q)\)) is
\begin{eqnarray}
\mathcal{L}_{\rm bdry}^{\partial Q}=(K-\Sigma)-\frac{\gamma}{4}\,\nabla_\mu\!\left[\phi\,\mathcal{G}^\mu\right],   
\end{eqnarray}
It is also useful to express the \(\gamma\)-completion as a total divergence intrinsic to \(\partial Q\), \(\nabla_\mu \mathcal{G}^\mu\), for an appropriately chosen boundary vector \(\mathcal{G}^\mu\) constructed from \(n^\mu\), \(K_{\mu\nu}\), \(K\), and \(\nabla_\mu\phi\). In this case, the cancelation mechanism becomes transparent: expanding the divergence reproduces the quadratic \((\nabla\phi)^2\)-weighted extrinsic-curvature structures needed to eliminate normal-derivative metric variations, while leaving the bulk equations unaffected. With these boundary terms included, the boundary metric variation yields the Neumann condition,
\begin{eqnarray}
K_{\alpha\beta}-h_{\alpha\beta}(K-\Sigma)+\mathcal{H}_{\alpha\beta}={\cal S}^{\partial\mathcal{Q}}_{\alpha\beta},    
\end{eqnarray}
while the scalar sector yields the bulk equation \(\nabla^\mu J_\mu=0\) (where we use the conditon ${\cal E}^{\mathcal{Q}}_{\phi}=0$) together with a boundary equation ensuring consistency between bulk and boundary scalar dynamics. In this case, we have that ${\cal F}^{\partial\mathcal{Q}}_{\phi}$ admits a first integral, leading to \({\cal F}^{\partial\mathcal{Q}}_{\phi} = {\cal G}_{0}\) which leads to \({\cal G}_{\mu\nu} = {\cal G}_{0}\), and considering \({\cal G}_0 = 0\), we have \({\cal G}_{\mu\nu} = 0\), which satisfies \({\cal G}_{\mu\nu} = 0\) and consequently leads to $\mathcal{H}_{\alpha\beta}=0$. These conditions lead to $\nabla^{\alpha}{\cal S}^{\partial\mathcal{Q}}_{\alpha\beta}=0$, which is the energy conservation \cite{Takayanagi:2011zk}. 

This boundary completion is closely tied to the causal/characteristic subtleties of Horndeski theories. In general, different perturbative modes need not propagate on the null cone of \(g_{\mu\nu}\); rather, their principal symbols define effective characteristic cones, and the domain of dependence is controlled by the outermost (fastest) one. The hyperbolicity of Horndeski hyperbolicity and effective metrics (see, e.g., \cite{Kovacs:2020pns}), ensuring a consistent initial–boundary value problem involves two logically distinct requirements: (i) the action must admit a differentiable variational principle under the chosen boundary data (addressed here by the \(\gamma\)-dependent boundary completion), and (ii) the resulting field equations must be hyperbolic in the regime of interest so that the causal development—and hence constructs like the Wheeler–DeWitt patch. In the symmetric “Einstein-like” backgrounds we focus on, the scalar principal operator is governed by \(Z^{\mu\nu}=a g^{\mu\nu}-\gamma G^{\mu\nu}\), and when \(G_{\mu\nu}\propto g_{\mu\nu}\) the corresponding effective cone becomes conformal to the metric cone, justifying the use of null hypersurfaces of \(g_{\mu\nu}\).

The subtlety relevant for the WdW patch is that the domain of dependence in a Horndeski theory is determined by the fastest characteristic cone among all propagating modes \cite{Kovacs:2020pns,Papallo:2017qvl,Bettoni:2016mij,Bruyere:2026gnt}. For the scalar degree of freedom, the principal (highest-derivative) part of the linearized equation around a background \((\bar g_{\mu\nu},\bar\phi)\) is controlled by \(Z^{\mu\nu}\). In the geometric-optics/eikonal limit \cite{Bruyere:2026gnt},

\begin{eqnarray}
\varphi(x)\sim e^{iS(x)/\epsilon},\qquad k_\mu=\nabla_\mu S,   
\end{eqnarray}
the leading-order scalar characteristic equation is
\begin{eqnarray}
Z^{\mu\nu}k_\mu k_\nu =0.   
\end{eqnarray}
This motivates defining an effective inverse metric for the scalar mode (up to an irrelevant positive conformal factor)
\begin{eqnarray}
(g_{\rm eff}^{-1})^{\mu\nu}\;\propto\;Z^{\mu\nu}=\alpha g^{\mu\nu}-\gamma G^{\mu\nu}.    
\end{eqnarray}
Hyperbolicity and absence of a wrong-sign kinetic term require \(Z^{\mu\nu}\) to be Lorentzian with the correct time orientation; a sufficient condition in the backgrounds we consider is that the effective kinetic coefficient
\begin{eqnarray}
\alpha_{\rm eff}\;\equiv\;\alpha+\gamma\Lambda_{\rm eff}    
\end{eqnarray}
remains positive. In the symmetric regimes employed in Secs. \ref{BHSL}-\ref{TEST} (planar AdS black hole / BTZ-type Einstein geometries, and the corresponding near-horizon and near-boundary regions used to construct the WdW patch), the background is (locally) Einstein,
\begin{eqnarray}
\bar G_{\mu\nu}=-\Lambda_{\rm eff}\,\bar g_{\mu\nu}.    
\end{eqnarray}
Then the scalar effective metric becomes conformal to the physical metric:
\begin{eqnarray}
Z^{\mu\nu}=\big(\alpha+\gamma\Lambda_{\rm eff}\big)\,\bar g^{\mu\nu}
\qquad\Longrightarrow\qquad
Z^{\mu\nu}k_\mu k_\nu=0 \;\Longleftrightarrow\; \bar g^{\mu\nu}k_\mu k_\nu=0,    
\end{eqnarray}
provided \(\alpha+\gamma\Lambda_{\rm eff}\neq 0\). Hence, in this regime, the scalar characteristic cone coincides with the null cone of the background metric, so the fastest scalar propagation is “null” with respect to the same null hypersurfaces used in Einstein gravity Fig \ref{fig:horndeski_cones_compact} \cite{Kovacs:2020pns}.


\begin{figure}[t]
\centering
\begin{tikzpicture}[scale=0.90, every node/.style={transform shape}]

\begin{scope}[shift={(-4.2,0)}]

  \node[font=\bfseries\small] at (0,3.8) {Bulk };

  \node[blue, anchor=east, font=\scriptsize] at (-2.6,1.2) 
    {$g^{\mu\nu}$};
  \node[red, anchor=east, font=\scriptsize] at (-2.6,0.4) 
    {$Z^{\mu\nu}$};
  \node[green!60!black, anchor=east, font=\scriptsize] at (-2.6,-0.4) 
    {$(g_{\rm fast}^{-1})^{\mu\nu}$};

  \fill[green!15,opacity=0.8] 
    (0,0) -- (2.1,2.1) arc[start angle=45, end angle=135, radius=2.97] 
    -- (-2.1,2.1) -- cycle;
  \fill[green!15,opacity=0.8] 
    (0,0) -- (2.1,-2.1) arc[start angle=-45, end angle=-135, radius=2.97] 
    -- (-2.1,-2.1) -- cycle;
  \draw[green!60!black,thick] (0,0) -- (2.1,2.1) (0,0) -- (-2.1,2.1);
  \draw[green!60!black,thick] (0,0) -- (2.1,-2.1) (0,0) -- (-2.1,-2.1);

  \fill[blue!15,opacity=0.8] 
    (0,0) -- (1.5,2.1) arc[start angle=54.5, end angle=125.5, radius=2.58] 
    -- (-1.5,2.1) -- cycle;
  \fill[blue!15,opacity=0.8] 
    (0,0) -- (1.5,-2.1) arc[start angle=-54.5, end angle=-125.5, radius=2.58] 
    -- (-1.5,-2.1) -- cycle;
  \draw[blue,thick] (0,0) -- (1.5,2.1) (0,0) -- (-1.5,2.1);
  \draw[blue,thick] (0,0) -- (1.5,-2.1) (0,0) -- (-1.5,-2.1);

  \fill[red!15,opacity=0.8] 
    (0,0) -- (1.0,2.1) arc[start angle=64.5, end angle=115.5, radius=2.32] 
    -- (-1.0,2.1) -- cycle;
  \fill[red!15,opacity=0.8] 
    (0,0) -- (1.0,-2.1) arc[start angle=-64.5, end angle=-115.5, radius=2.32] 
    -- (-1.0,-2.1) -- cycle;
  \draw[red,thick] (0,0) -- (1.0,2.1) (0,0) -- (-1.0,2.1);
  \draw[red,thick] (0,0) -- (1.0,-2.1) (0,0) -- (-1.0,-2.1);

  \node[align=center, font=\tiny] (domtext) at (0,-3.0) {Domain of dependence};
  \draw[->,thin] (domtext.north) -- (1.2,-2.0);

  \node[font=\small\bfseries] at (0,-3.6) {(a)};

\end{scope}

\begin{scope}[shift={(4.2,0)}]

  \node[font=\bfseries\small] at (0,3.8) {Boundary};

  \node[blue, anchor=east, font=\scriptsize] at (-2.6,0.8) 
    {$h_{\alpha\beta}$};
  \node[green!60!black, anchor=east, font=\scriptsize] at (-2.6,0.0) 
    {$(h_{\rm eff}^{-1})^{\alpha\beta}$};

  \draw[black, line width=1.5pt] (0,-2.3) -- (0,2.3);
  \node[anchor=south,black,font=\tiny\bfseries] at (0,2.4) {AdS boundary};

  \fill[green!15,opacity=0.8] (0,0) -- (1.8,2.1) -- (-1.8,2.1) -- cycle;
  \fill[green!15,opacity=0.8] (0,0) -- (1.8,-2.1) -- (-1.8,-2.1) -- cycle;
  \draw[green!60!black,thick] (0,0) -- (1.8,2.1) (0,0) -- (-1.8,2.1);
  \draw[green!60!black,thick] (0,0) -- (1.8,-2.1) (0,0) -- (-1.8,-2.1);

  \fill[blue!15,opacity=0.8] (0,0) -- (1.2,2.1) -- (-1.2,2.1) -- cycle;
  \fill[blue!15,opacity=0.8] (0,0) -- (1.2,-2.1) -- (-1.2,-2.1) -- cycle;
  \draw[blue,thick] (0,0) -- (1.2,2.1) (0,0) -- (-1.2,2.1);
  \draw[blue,thick] (0,0) -- (1.2,-2.1) (0,0) -- (-1.2,-2.1);

  \node[font=\small\bfseries] at (0,-3.6) {(b)};

\end{scope}

\end{tikzpicture}
\caption{Cones in Horndeski AdS/BCFT. (a) Bulk cotangent space with physical metric $g^{\mu\nu}$, scalar effective metric $Z^{\mu\nu}$, and fastest cone $(g_{\rm fast}^{-1})^{\mu\nu}$. (b) Boundary/brane tangent space with induced metric $h_{\alpha\beta}$ and effective cone $(h_{\rm eff}^{-1})^{\alpha\beta}$. Hence, in the parameter regime and geometries we actually use, the fastest characteristic cone coincides with the standard null cone, and the WdW patch is correctly defined by null hypersurfaces of \(\bar g_{\mu\nu}\).}
\label{fig:horndeski_cones_compact}
\end{figure}


A physically motivated parameter restriction that ensures the above cone is well-defined and avoids pathological superluminal widening in the examples we study is
\begin{eqnarray}
\alpha_{\rm eff}=\alpha+\gamma\Lambda_{\rm eff}>0,
\qquad\text{and we focus on}\qquad \gamma\le 0,    
\end{eqnarray}
which is also consistent with the causality considerations discussed in the Horndeski/braneworld literature (e.g. \cite{Minamitsuji:2013vra}) and with our numerical choices in Sec. \ref{NU}. Under these assumptions, the null hypersurfaces of \(\bar g_{\mu\nu}\) provide the relevant boundary of the domain of dependence that defines the WdW patch, justifying our use of null rays in the Penrose/Kruskal constructions in Secs. \ref{BHSL}-\ref{TEST}.

Outside these symmetric/Einstein regimes (or for other parameter ranges), different modes can propagate on different characteristic cones \cite{Kovacs:2020pns,Papallo:2017qvl,Bettoni:2016mij}, and the WdW patch should, in principle, be defined using the outermost characteristic hypersurface. Our analysis is explicitly restricted to the class of backgrounds and couplings for which the effective cone relevant to the fastest mode is captured by the null structure employed in our calculations.

\section{Derivation of Entropy in Horndeski with BCFT contributions}\label{WEHT} 
 In this section we summarize and slightly reorganize known results on black hole entropy in Horndeski gravity, following in particular. Our purpose is to collect these formulas in a form adapted to the AdS/BCFT setting of this paper, rather than to present new derivations. As we know, the temperature and entropy of black holes with charges and angular momentum \cite{Feng:2015oea,Hajian:2020dcq,Feng:2015wvb} are not readily amenable to the usual techniques, such as holographic renormalization, for computing their entropy \cite{Santos:2024cvx}. Specially for a general covariant gravitational theory, where the Lagrangian \(\mathcal{L}^{{\cal Q}}_{H}\) depends on the spacetime metric \(g_{\mu\nu}\), the Riemann tensor \(R_{\mu\nu\alpha\beta}\), its covariant derivatives, the coupling between the kinetic term and the Einstein tensor as \(G^{\mu\nu}\nabla_\mu\phi\nabla_\nu\phi\), and second-order terms. In this section, we present the Wald entropy formula (\ref{W1}) and the Wald formalism (\ref{W2}), which are fundamental to guide the construction of the holographic renormalization scheme (\ref{W3}) to satisfy the first law of thermodynamics with boundary contributions (BCFT). 

\subsection{Wald entropy formula}\label{W1}
The Wald entropy formula is given by \cite{Feng:2015oea,Hajian:2020dcq,Feng:2015wvb,Tsilioukas:2024seh,Tsilioukas:2025dmy}:
\begin{eqnarray}
{\cal S}^{{\cal\,Q}}_{W}=\int_{\mathcal{H}}{\sqrt{|\eta|}\,\mathbf{X}^{\sigma\rho}_{{\cal Q}}\boldsymbol{\epsilon}_{\sigma\rho}},   
\end{eqnarray}
where \(\mathcal{H}\) is the black hole horizon, \(\boldsymbol{\epsilon}_{\sigma\rho}\) is the binormal tensor \cite{Hajian:2020dcq}, and \(\mathbf{X}^{\sigma\rho}_{{\cal Q}}\) is defined as:\begin{eqnarray}
\mathbf{X}^{\sigma\rho}_{{\cal\,Q}}=-\frac{\delta \mathcal{L}^{{\cal Q}}_H}{\delta R_{\mu\nu\sigma\rho}} \boldsymbol{\epsilon}_{\sigma\rho},    
\end{eqnarray}
with
\begin{eqnarray}
&&\,\,\,\,\,\,\,\,\,\,\,\,\,\,\,\,\,\,\,\,\,\,\,\,\,\,\,\,\,\,\,\,\,\,\,\,\,\,\,\,\,\,\,\,\,\,\,\,\,\,\,\,\,\,\,\,\,\,\,\,\,\frac{\partial{\cal L}^{{\cal Q}}_H}{\partial R_{\mu\nu\rho\sigma}}=-\frac{1}{16\pi\,G_{N}}(g^{\mu\rho}g^{\nu\sigma} - g^{\nu\rho}g^{\mu\sigma})\nonumber\\
&&\,\,\,\,\,\,\,\,\,\,\,\,\,\,\,\,\,\,\,\,\,\,\,\,\,\,\,\,\,\,\,\,\,\,\,\,\,\,\,\,\,\,\,\,\,\,\,\,\,\,\,\,\,\,\,\,\,\,\,\,\,-\frac{1}{16\pi\,G_{N}}\frac{\gamma\,G_{N}}{4}[ g^{\mu\rho}\phi^{,\nu}\phi^{,\sigma} - g^{\nu\rho}\phi^{,\mu}\phi^{,\sigma} + g^{\nu\sigma}\phi^{,\mu}\phi^{,\rho} \nonumber \\
&&\,\,\,\,\,\,\,\,\,\,\,\,\,\,\,\,\,\,\,\,\,\,\,\,\,\,\,\,\,\,\,\,\,\,\,\,\,\,\,\,\,\,\,\,\,\,\,\,\,\,\,\,\,\,\,\,\,\,\,\,\,-(g^{\mu\sigma}\phi^{,\nu}\phi^{,\rho}g^{\mu\rho}g^{\nu\sigma} - g^{\nu\rho}g^{\mu\sigma}) \phi^{,\lambda} \phi_{,\lambda}]\label{dLdR}   
\end{eqnarray}
where $\phi^{,\alpha} = g^{\alpha\beta} \phi_{,\beta}$. For a metric expressed in complex coordinates as:
\begin{eqnarray}
ds^2 = dz d\bar{z} + \eta_{ij} dy^i dy^j.   
\end{eqnarray}
Replacing the components of this metric into the expression (\ref{dLdR}) yields:
\begin{eqnarray}
\frac{\partial\mathcal{L}^{{\cal Q}}_H}{\partial R_{z\bar{z}z\bar{z}}} =\frac{1}{16\pi\,G_{N}}\left[1 - \frac{\gamma\,G_{N}}{4}\left( \phi^{,\lambda} \phi_{,\lambda} - \phi^{,z} \phi^{,\bar{z}} \right)\right].    
\end{eqnarray}
Expanding the term \(\phi^{,\lambda} \phi_{,\lambda}\) to \(\phi^{,\lambda} \phi_{,\lambda}=\phi^{,z} \phi^{,\bar{z}} + \eta_{ij} \phi^{,i} \phi^{,j}\), the above expression further simplifies to:
\begin{eqnarray}
\frac{\partial\mathcal{L}^{{\cal Q}}_H}{\partial R_{z\bar{z}z\bar{z}}}=\frac{1}{16\pi\,G_{N}}\left[1-\frac{\gamma\,G_{N}}{4} \eta_{ij} \phi^{,i} \phi^{,j}\right].    
\end{eqnarray}
Consequently, the functional form of the entropy becomes \cite{Santos:2024cvx,Caceres:2017lbr,Feng:2015oea}:
\begin{eqnarray}
{\cal S}^{{\cal\,Q}}_{W}= \int_{{\cal\,Q}}{\sqrt{|\eta|} \frac{\partial \mathcal{L}^{{\cal Q}}_H}{\partial R_{z\bar{z}z\bar{z}}}}= \frac{1}{16\pi\,G_{N}}\int_{{\cal\,Q}}{\sqrt{|\eta|}\left[1-\frac{\gamma\,G_{N}}{4} \eta_{ij} \phi^{,i} \phi^{,j} \right]}. \label{EEG}    
\end{eqnarray}
As a result, the entropy includes a Wald-like correction that is proportional to the squared norm of the scalar field's gradient on the surface. Now, following the prescription of \cite{Minamitsuji:2013vra} we can write $\mathcal{L}^{{\cal\partial\,Q}}_{bdry}$
\begin{eqnarray}
\mathcal{L}^{{\cal\partial\,Q}}_{bdry}=\frac{1}{16\pi\,G_{N}}(K-\Sigma)-\frac{1}{16\pi\,G_{N}}\frac{\gamma\,G_{N}}{4}[(\nabla_\mu\phi\nabla_\nu\phi\,n^\mu\,n^\nu-(\nabla\phi)^2)K-\nabla_\mu\phi\nabla_\nu\phi\,K^{\mu\nu}],\nonumber    
\end{eqnarray}
as 
\begin{eqnarray}
&&\mathcal{L}^{{\cal\partial\,Q}}_{bdry}=\frac{1}{16\pi\,G_{N}}(K-\Sigma)-\frac{1}{16\pi\,G_{N}}\frac{\gamma\,G_{N}}{4}\nabla_\mu\phi\nabla_\nu\phi[(g_{\mu\nu}n^\mu\,n^\nu\,g_{\mu\nu}-\,n^\mu\,n^\nu\,g_{\mu\nu})h_{\mu\nu}K-K_{\mu\nu}],\nonumber\\
&&\mathcal{L}^{{\cal\partial\,Q}}_{bdry}=\frac{1}{16\pi\,G_{N}}(K-\Sigma)+\frac{1}{16\pi\,G_{N}}\frac{\gamma\,G_{N}}{4}\nabla_\mu\phi\nabla_\nu\phi(K_{\mu\nu}-h_{\mu\nu}K).\nonumber
\end{eqnarray}
Deriving $\mathcal{L}^{{\cal\partial\,Q}}_{bdry}$ with respect to $y$, we have the following equation:
\begin{eqnarray}
\frac{\partial}{\partial\,y}\mathcal{L}^{{\cal\partial\,Q}}_{bdry}=\frac{1}{16\pi\,G_{N}}K_{,y}+\frac{1}{16\pi\,G_{N}}\frac{\gamma\,G_{N}}{4}\nabla_\mu\phi\nabla_\nu\phi(K^{,y}_{\mu\nu}-h_{\mu\nu}K^{,y}),
\end{eqnarray}
where the gravitational equation contains the second-order derivatives, such as
\begin{eqnarray}
K_{\mu\nu,y} - h_{\mu\nu} K_{,y} = -G_{y\nu} + \cdots. 
\end{eqnarray}
Now, we can write
\begin{eqnarray}
&&\mathcal{L}^{{\cal\partial\,Q}}_{,y}=\frac{1}{16\pi\,G_{N}}(K^{\rho}_{\,\,\mu|\rho}-G_{\rho\nu} + \cdots)\cr
&&\,\,\,\,\,\,\,\,\,\,\,\,\,\,\,+\frac{1}{16\pi\,G_{N}}\frac{\gamma\,G_{N}}{4}\left(R^{\rho}_{\mu\rho\nu}-\frac{1}{2}g_{\mu\nu}g^{\alpha\beta}R^{\rho}_{\alpha\rho\beta}\,+...\right)\nabla_\mu\phi\nabla_\nu\phi.
\end{eqnarray}
Here, $\mathcal{L}^{{\cal\partial\,Q}}_{,y}$ denotes the induced metric contributions to the bulk and is in agreement with the descriptions of \cite{Caceres:2017lbr,Dong:2013qoa}. As we know, the terms of extrinsic curvature in $\mathcal{L}^{{\cal\partial\,Q}}_{,y}$ vanish in the variation of the lagrangian concerning the Riemann tensor. So, we have
\begin{eqnarray}
\frac{\partial\mathcal{L}^{{\cal\partial\,Q}}_{,y}}{\partial\,R_{\mu\nu\rho\sigma}}=\frac{1}{16\pi\,G_{N}}\left[1-\frac{\gamma\,G_{N}}{4}h_{ij}\phi^{,i} \phi^{,j}\right],   
\end{eqnarray}
The functional form of the entropy boundary is given by 
\begin{eqnarray}
{\cal S}^{{\partial\cal\,Q}}_{W}= \int_{\partial{\cal\,Q}}{\sqrt{|\eta|}\frac{\partial\mathcal{L}^{{\cal\partial\,Q}}_{,y}}{\partial\,R_{\mu\nu\rho\sigma}}}= \frac{1}{8\pi\,G_{N}}\int_{\partial{\cal\,Q}}{\sqrt{|h|}\left[1-\frac{\gamma\,G_{N}}{4}h_{ij} \phi^{,i} \phi^{,j} \right]}. \label{EEG1}  
\end{eqnarray}
Now, we can conclude that ${\cal S}_{W}={\cal S}^{{\cal\,Q}}_{W}+{\cal S}^{{\partial\cal\,Q}}_{W}$, that is,
\begin{eqnarray}
&&{\cal S}_{W}=\frac{1}{16\pi\,G_{N}}\int_{{\cal\,Q}}{\sqrt{|\eta|}\left[1-\frac{\gamma\,G_{N}}{4} \eta_{ij} \phi^{,i} \phi^{,j} \right]}\nonumber\\
&&\,\,\,\,\,\,\,\,\,\,\,\,\,\,\,\,\,\,\,\,\,\,\,\,\,\,\,\,\,\,\,\,\,\,\,\,\,\,\,\,\,\,\,\,\,\,\,\,+\frac{1}{8\pi\,G_{N}}\int_{\partial{\cal\,Q}}{\sqrt{|h|}\left[1-\frac{\gamma\,G_{N}}{4}h_{ij} \phi^{,i} \phi^{,j} \right]}.\label{EEG2}    
\end{eqnarray}
This result agrees with the universal description of \cite{Caceres:2017lbr,Dong:2013qoa}, where the expression is used to calculate the entropy in theories that are dual to second derivative gravity, where the Lagrangian is constructed as a contraction of the Riemann tensors. This Wald's entropy formula for black holes through ${\cal L}^{{\cal\,Q}}_{H}$, along with additional terms that account for corrections with the extrinsic curvature through ${\cal L}^{\partial{\cal\,Q}}_{bdry}$. 

\subsection{Wald formalism}\label{W2}

As discussed in \cite{Santos:2024cvx,Caceres:2017lbr,Feng:2015oea}, let us revisit the black hole entropy using the Wald formalism. This approach provides two key results regarding entropy: 
\begin{itemize}
\item The first law of black hole mechanics.
\item A relation that connects the entropy to the integral of the Noether charge (associated with diffeomorphism invariance) over the black hole's bifurcation surface.
\end{itemize}
To derive the first result, the first law, we evaluate a closed differential form given by \(\delta \mathbf{Q} - \xi \cdot \mathbf{\Theta}\). Here, \(\delta \mathbf{Q}\) represents the on-shell variation of the Noether charge (corresponding to diffeomorphism invariance), \(\xi\) is the timelike Killing vector field that becomes null on the bifurcation surface, and \(\Theta\) is the boundary term arising from the variation of the gravitational action. In Horndeski theory, the variation of the fields in the action $\mathbf{{\cal\,L}}^{{\cal Q}}_{H}$ leads to a surface term \(J^\mu_{{\cal\,Q}}\), which is expressed as:
\begin{eqnarray}
J^\mu_{{\cal\,Q}} = 2 \frac{\partial\mathbf{{\cal\,L}}^{{\cal Q}}_{H}}{\partial R_{\rho\mu\nu\sigma}} \delta g_{\rho\sigma} + \frac{\partial \mathbf{{\cal\,L}}^{{\cal Q}}_{H}}{\partial R_{\rho\sigma\mu\nu}} \nabla_\sigma \delta g_{\rho\nu} - 2 \nabla_\nu \left( \frac{\partial\mathbf{{\cal\,L}}^{{\cal Q}}_{H}}{\partial (\nabla_\mu \phi)} \delta \phi \right),    
\end{eqnarray}
where $\mathbf{{\cal\,L}}^{{\cal Q}}_{H}$ is the Lagrangian density. This term can be decomposed into contributions from different components of the theory:
\begin{eqnarray}
J^\mu_{{\cal\,Q}} = \kappa J^\mu_{{\cal\,Q},g} + \alpha J^\mu_{{\cal\,Q},\phi} + \gamma (J^\mu_{gc} + J^\mu_{\phi c}).    
\end{eqnarray}
With the individual terms defined as:
\begin{itemize}
\item \(J^\mu_{{\cal\,Q},g} = g^{\mu\rho} g^{\nu\sigma} (\nabla_\sigma \delta g_{\nu\rho} - \nabla_\rho \delta g_{\nu\sigma})\),
\item \(J^\mu_{{\cal\,Q},\phi} = -g^{\mu\nu} \nabla_\nu \phi \delta \phi\),
\item \(J^\mu_{\phi\,,c} = G^{\mu\nu} \nabla_\nu \phi \delta \phi\),
\item \(J^\mu_{gc}\) includes additional terms involving derivatives of \(\phi\) and the metric.
\end{itemize}
The solution phase space method is used. For a general \(n\)-dimensional covariant theory with Lagrangian \(\mathcal{L}^{{\cal Q}}_{H}\), the dynamical fields are collectively denoted by \(\Phi\), and their solutions by \(\bar{\Phi}\) \cite{Iyer:1994ys,Barnich:2007bf}. The Noether current \((n-1)\)-form associated with a vector field \(\xi^\mu\) is:
\begin{eqnarray}
\mathbf{J}_\xi = \mathbf{\Theta}(\delta_\xi \Phi) - \xi \cdot \mathbf{{\cal\,L}}^{{\cal Q}}_{H},    
\end{eqnarray}
where \(\mathbf{\Theta}\) is derived from the Lagrangian variation, \(\delta \mathbf{{\cal\,L}}^{{\cal Q}}_{H} =\mathbf{{\cal\,E}}^{{\cal\,Q}}_{\mu\nu}\delta \Phi + d\mathbf{\Theta}(\delta \Phi)\), and \(\mathbf{{\cal\,E}}^{{\cal\,Q}}_{\mu\nu}\) represents the equations of motion \cite{Barnich:2007bf}. On-shell, \(\mathbf{J}_\xi\) becomes exact, that is, \(\mathbf{J}_\xi = d\mathbf{Q}^{{\cal\,Q}}_\xi\), where \(\mathbf{Q}^{{\cal\,Q}}_\xi\) is the Noether charge density. $\delta H_{\xi_H}=\int_{\text{H}} \boldsymbol{k}_\xi(\delta\Phi, \bar\Phi).$ The entropy variation $\delta\,{\cal S}_{W}$ that satisfies a consistent first law is then defined as $\delta H_{\xi_H}:={\cal T}\delta\,{\cal S}^{{\cal Q}}_{Wald}$, that is, the entropy variation is defined as
\begin{eqnarray}
{\cal T}{\cal S}^{{\cal Q}}_{W}=\int_{\mathcal{H}} \boldsymbol{k}_{\xi_H}(\delta\Phi, \bar{\Phi}),    
\end{eqnarray}
where \(\boldsymbol{k}_{\xi_H} = \delta \mathbf{Q}^{{\cal\,Q}}_{\xi_H} - \xi_H \cdot \mathbf{\Theta}(\delta \Phi)\), and \({\cal T}\) is the black hole temperature. This approach ensures a consistent application of the first law of black hole thermodynamics \cite{Hajian:2020dcq}. The variation of the entropy, including the boundary terms, is as follows:
\begin{eqnarray}
{\cal T}{\cal S}_{W}={\cal T}(\delta {\cal S}^{{\cal Q}}_{W}+\delta {\cal S}^{{\cal\partial\,Q}}_{W})=\int_{\mathcal{H}} \left( \delta \mathbf{Q}^{{\cal\,Q}}_{\xi_H} - \xi_H \cdot \mathbf{\Theta}(\delta \Phi) + \delta \mathbf{Q}_{\xi_H}^{{\partial\,Q}} \right),    
\end{eqnarray}
where \(\delta \mathbf{Q}^{{\partial\,Q}}_{\xi_H}\) is the variation of the Noether charge associated with the boundary Lagrangian. So, for the boundary, we have
\begin{eqnarray}
J^\mu_{{\cal\partial\,Q}} =-2\nabla_\nu \left( \frac{\partial\mathbf{{\cal\,L}}^{{\cal\partial\,Q}}_{H}}{\partial (\nabla_\mu \phi)} \delta \phi \right)\,\,;\,\,\delta\mathbf{Q}^{{\partial\,Q}}_{\xi_H}= \frac{\partial\mathbf{{\cal\,L}}^{{\cal\partial\,Q}}_{H}}{\partial (\nabla_\mu \phi)} \delta \phi.    
\end{eqnarray}

\subsection{Black hole solution}
We now provide a black hole solution to apply the Wald formalism presented in the previous section (\ref{W2}). In Horndeski gravity, black hole solutions are described by the following metric:
\begin{eqnarray}
ds^2 = -h(r) dt^2 + \frac{dr^2}{f(r)} + r^2 d\Omega_{d-2,\epsilon}^2,\label{StationaryBHAnsatz}  
\end{eqnarray}
where \(d\) represents the spacetime dimension, and \(\epsilon = -1, 0, 1\) corresponds to hyperbolic, planar, and spherical horizon geometries, respectively. The metric functions \(h(r)\) and \(f(r)\) are solutions of equation ${\cal E}^{\mathcal{Q}}_{\mu\nu}=0$ with the following form \cite{Caceres:2017lbr,Feng:2015oea}:
\begin{eqnarray}
&&h(r) = \frac{(d-1)^2 \beta^2 \gamma^2 g^4 r^4}{\epsilon (d+1)(d-3)(4\kappa + \beta\gamma)^2} \, {}_2F_1\left(1, \frac{1}{2}(d+1), \frac{1}{2}(d+3), -\frac{d-1}{(d-3)\epsilon} g^2 r^2\right)\cr
&&- \frac{\mu}{r^{d-3}} + \frac{8\kappa[g^2 r^2 (2\kappa + \beta\gamma) + 2\epsilon\kappa]}{(4\kappa + \beta\gamma)^2},\\
&&f(r) = \frac{(4\kappa + \beta\gamma)^2 \left[(d-1)g^2 r^2 + (d-3)\epsilon\right]^2}{\left[(d-1)(4\kappa + \beta\gamma)g^2 r^2 + 4(d-3)\epsilon\kappa\right]^2} h(r).
\end{eqnarray}
The scalar field \(\phi(r)\) satisfies the following equation ${\cal E}^{\mathcal{Q}}_{\phi}=0$:
\begin{eqnarray}
\left(\frac{d\phi}{dr}\right)^2 = \frac{\beta}{f(r)} \left(1 + \frac{(d-3)\epsilon}{(d-1)g^2 r^2}\right)^{-1}\label{dchidr}.    
\end{eqnarray}
Here, \(g\) and \(\beta\) are constants related to the couplings \(\alpha\) and \(\Lambda\) in the action (\ref{1}). These relationships are expressed as follows:
\begin{eqnarray}
&&\alpha = \frac{1}{2}(d-1)(d-2)g^2\gamma,\\
&&\Lambda = -\frac{1}{2}(d-1)(d-2)g^2 \left(1 + \frac{\beta\gamma}{2\kappa}\right).
\end{eqnarray}
The integral of the Noether charge \(\mathbf{Q}\) on the bifurcation surface is equal to the product of the temperature and entropy of the black hole. Let us now explore how this relationship is modified in the context of Horndeski gravity. For a stationary black hole metric of the form given in (\ref{StationaryBHAnsatz}), the Noether charge was derived in \cite{Feng:2015oea}. It can be expressed as \(\mathbf{Q} = \mathbf{Q}^{{\cal\,Q}}_{\text{Einstein}} + \mathbf{Q}^{{\cal\,Q}}_{\gamma}+\mathbf{Q}^{{\cal\partial\,Q}}_{\gamma}\), where \(\mathbf{Q}^{{\cal\,Q}}_{\text{Einstein}}\) represents the contribution from Einstein gravity with minimal coupling, and \(\mathbf{Q}^{{\cal\,Q}}_{\gamma}+\mathbf{Q}^{{\cal\partial\,Q}}_{\gamma}\) accounts for the non-minimal coupling effects. These components are given by:
\begin{eqnarray}
&&\mathbf{Q}^{{\cal\,Q}}_{\text{Einstein}} = \frac{\Delta_{{\cal Q}}r^{d-2}}{16\pi G_{N}} \sqrt{\frac{f}{h}} h' \Omega_{d-2},\\  
&&\mathbf{Q}^{{\cal\,Q}}_{\gamma} = -\frac{\Delta_{{\cal Q}}}{32\pi\,G_{N}}(d-2)\gamma r^{d-3} \sqrt{\frac{h}{f}} f^2 (\phi')^2 \Omega_{d-2},\\
\end{eqnarray}
 where the construction of $\mathbf{Q}^{{\cal\partial\,Q}}_{\gamma}$ is based on the fact that ${\cal F}^{\partial\mathcal{Q}}_{\phi}$ have a first integral, expressed as \({\cal F}^{\partial\mathcal{Q}}_{\phi} = {\cal G}_{0}\). Consequently, we have \({\cal G}_{\mu\nu} = {\cal G}_{0}\), where \({\cal G}_0\) is a constant. For simplicity, we set \({\cal G}_0 = 0\), which leads to \({\cal G}_{\mu\nu} = 0\). This condition implies that \(K_{\mu\nu} - h_{\mu\nu}K = 0\), which satisfies \({\cal G}_{\mu\nu} = 0\), and with this we can find
\begin{eqnarray}
&&\mathbf{Q}^{{\cal\partial\,Q}}_{\gamma} = -\frac{\Delta_{{\cal\partial\,Q}}}{16\pi\,G_{N}}(d-2)\gamma r^{d-3} \sqrt{\frac{h}{f}} f^2 (\phi')^2 \Omega_{d-2}.
\end{eqnarray}
In the CFT framework, \(\Delta_{{\cal\partial\,Q}}\) represents the "width" of the boundary; \(\Delta_{{\cal\,Q}}\) is the "width" of the bulk. From these expressions, it is evident that \(\mathbf{Q}^{{\cal\,Q}}_{\gamma}+\mathbf{Q}^{{\cal\partial\,Q}}_{\gamma}\) vanishes at the horizon. This can be shown by replacing \((\phi')^2\) in (\ref{dchidr}) and noting that the ratio \(h/f\) remains regular on the horizon. As a result, the integral of \(\mathbf{Q}\) on the bifurcation surface simplifies the contribution of \(\mathbf{Q}^{{\cal\,Q}}_{\text{Einstein}}\) alone. This integral is equal to the product of temperature \({\cal\,T}\), and the entropy \({\cal S}_{W}\). For the planar case (\(\epsilon = 0\)), as discussed in \cite{Caceres:2017lbr, Feng:2015oea}, the following expressions are obtained:
\begin{eqnarray}
&&\int_{\infty} \phi = \frac{(d-2)}{16\pi G_{N}} \left( 1 + \frac{\gamma \beta G_{N}}{4} \right) \delta \mu,\\ 
&&\int_{\mathcal{H}} \phi =\left[\frac{\Delta_{{\cal Q}}(d-1)(d-2)g^2}{16\pi G_{N}} \left( 1 + \frac{\gamma \beta G_{N}}{4} \right) r_h^{d-2} \delta r_h\right]_{{\cal\,Q}},\\
&&\int_{\mathcal{H}} \phi =\left[\frac{\Delta_{{\cal\partial\,Q}}(d-1)(d-2)g^2}{8\pi G_{N}} \left( 1 + \frac{\gamma \beta G_{N}}{4} \right) r_h^{d-2} \delta r_h\right]_{{\cal\partial\,Q}}.
\end{eqnarray}
The terms with \(\gamma\) in these equations arise due to the singular behavior of the scalar field near the singularity. To address this issue, the authors of \cite{Caceres:2017lbr, Feng:2015oea} propose a straightforward approach: \(\delta {\cal M}\) is the variation of the mass and \({\cal T} \delta {\cal S}_{W}\) is the variation of entropy times the temperature. Using the above integrals, we have:
\begin{eqnarray}
&&{\cal M} = \frac{(d-2)}{16\pi G_{N}} \left( 1 + \frac{\gamma \beta G_{N}}{4} \right) \mu,\\ 
&&{\cal S}^{{\cal Q}}_{W} = \frac{\Delta_{{\cal Q}}}{4G_{N}} \left( 1 + \frac{\gamma \beta G_{N}}{4} \right) r_h^{d-2},\\
&&{\cal S}^{{\cal\partial\,Q}}_{W} = \frac{\Delta_{{\cal\partial\,Q}}}{G_{N}} \left( 1 + \frac{\gamma \beta G_{N}}{4} \right) r_h^{d-2},
\end{eqnarray}
where $\delta{\cal S}_{W}= \delta {\cal S}^{{\cal Q}}_{W}+\delta {\cal S}^{{\cal\partial Q}}_{W}$. These definitions have contributions of the non-minimal coupling parameter \(\gamma\) and ensure consistency with the planar black hole solution. The first law is automatically satisfied ${\cal T}\delta{\cal S}_{W}={\cal T}( \delta {\cal S}^{{\cal Q}}_{W}+\delta {\cal S}^{{\cal\partial Q}}_{W})=\delta {\cal M}$. This formula includes both bulk and boundary contributions, ensuring a complete description of black hole entropy in the presence of the boundary Lagrangian \cite{Khodabakhshi:2020fhb,Caceres:2017lbr}. 

\subsection{Holographic renormalization scheme}\label{W3}
We introduce the holographic renormalization scheme, which is used to compute the Euclidean on-shell action \cite{Santos:2024cvx}. This action is directly connected to the free energy of the corresponding thermodynamic system. Within the framework of the AdS/CFT correspondence \cite{Santos:2024cwf,Takayanagi:2011zk}, holographic renormalization is a systematic approach to eliminate divergences from infinite quantities on the gravitational side of duality \cite{Santos:2024zoh}. This process mirrors the standard renormalization techniques applied to gauge field theories on the boundary \cite{Santos:2021orr,Santos:2023flb,Feng:2015oea}.

Considering the Euclidean action, expressed as:
\begin{eqnarray}
 I_{\rm E} = I_{{\cal Q}} + 2I_{{\cal\partial\,Q}},   
\end{eqnarray}
where the bulk Euclidean action is given by:

\begin{eqnarray}
&&I_{{\cal\,Q}} =-\frac{1}{\kappa }\int_{\mathcal{Q}} d^dx \sqrt{g} \left[ (R - 2\Lambda) + \frac{\gamma}{2} G_{\mu\nu} \nabla^\mu \phi \nabla^\nu \phi \right] \\
&& - \frac{2}{ \kappa } \int_{\mathcal{M}} d^{d-1}x \sqrt{\bar{\gamma}} \left[ (K^{(\bar{\gamma})} - \Sigma^{(\bar{\gamma})}) - \frac{\gamma}{4} \left( \nabla_\mu \phi \nabla_\nu \phi n^\mu n^\nu - (\nabla \phi)^2 \right) K^{(\bar{\gamma})} - \frac{\gamma}{4} \nabla^\mu \phi \nabla^\nu \phi K^{(\bar{\gamma})}_{\mu\nu} \right].\nonumber
\end{eqnarray}
Here, \(g\) represents the determinant of the bulk metric \(g_{\mu\nu}\) in the manifold \(\mathcal{Q}\), while \(\bar{\gamma}\) is the induced metric on the boundary surface \(\mathcal{M}\), which has a tension \(\Sigma^{(\bar{\gamma})}\). The extrinsic curvature of the boundary is characterized by its trace \(K^{(\bar{\gamma})}\).

\begin{figure}[!ht]
\begin{center}
\includegraphics[width=\textwidth]{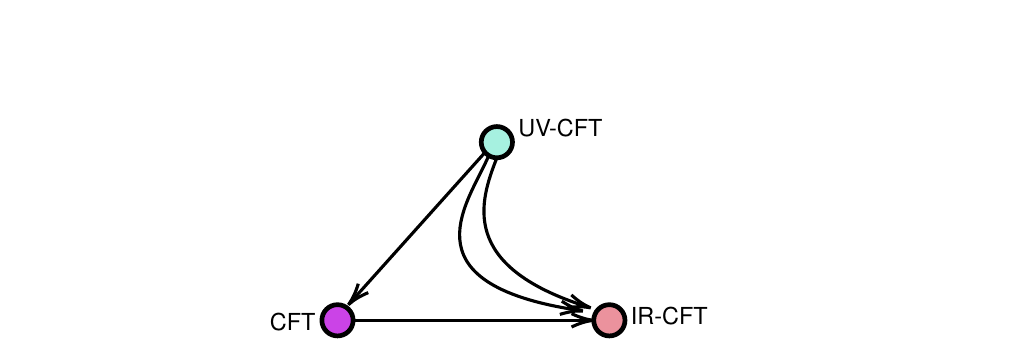}
\caption{Organized scheme of CFT space.}
\label{BTZ2}
\end{center}
\end{figure}

For the boundary contribution, the Euclidean action is expressed as:

\begin{eqnarray}
&&I_{\partial{\cal\,Q}} = -\frac{1}{ \kappa }\int_{\mathcal{Q}}{d^dx \sqrt{g} \left[ (R - 2\Lambda) + \frac{\gamma}{2} G_{\mu\nu} \nabla^\mu \phi \nabla^\nu \phi \right]} \\
&& - \frac{2}{ \kappa } \int_{\partial{\cal\,Q}}{ d^{d-1}x \sqrt{h} \left[ (K - \Sigma) - \frac{\gamma}{4} \left( \nabla_\mu \phi \nabla_\nu \phi n^\mu n^\nu - (\nabla \phi)^2 \right) K - \frac{\gamma}{4} \nabla^\mu \phi \nabla^\nu \phi K_{\mu\nu} \right]}.\nonumber
\end{eqnarray}
With \(h\) being the determinant of the boundary metric \(h_{\mu\nu}\) in the manifold \(\partial\mathcal{Q}\). In this formulation, the AdS/CFT correspondence reveals a deep connection between infrared (IR) divergences in the gravitational bulk and ultraviolet (UV) divergences in the boundary conformal field theory (CFT). This relationship, often referred to as the IR-UV connection, is shown in Fig. \ref{BTZ2}.

In order to achieve consistency between Wald's formula and its formalism (\ref{W1})-(\ref{W2}), we consider again that \({\cal F}^{\partial\mathcal{Q}}_{\phi}={\cal G}_{0}\) with \({\cal G}_{\mu\nu}={\cal G}_{0}\), we set \({\cal G}_0 = 0\) \footnote{However, for ${\cal G}_0\neq0$, we have the cases of \cite{Santos:2024zoh,Santos:2021orr,Santos:2023flb,Feng:2015oea}.}, and for the metric (\ref{StationaryBHAnsatz}), we have

\begin{eqnarray}
&&{\cal S}^{{\cal Q}}_{W} = \frac{\Delta_{{\cal Q}}}{4G_{N}} \left( 1 + \frac{\gamma \beta G_{N}}{4} \right) r_h^{d-2},\\
&&{\cal S}^{{\cal\partial\,Q}}_{W} = \frac{\Delta_{{\cal\partial\,Q}}}{G_{N}} \left( 1 + \frac{\gamma \beta G_{N}}{4} \right) r_h^{d-2}.
\end{eqnarray}    
where $\delta{\cal S}_{W}= \delta {\cal S}^{{\cal Q}}_{W}+\delta {\cal S}^{{\cal\partial Q}}_{W}$, with $\gamma$-correction for the Takayanagi prescription \cite{Santos:2024cwf,Takayanagi:2011zk,Santos:2024cvx}. In the constructions presented in subsections (\ref{W1}) and (\ref{W2}), we introduce the Wald entropy formula and the Wald formalism for computing entropy in the Horndeski framework. These two formalisms agree with the holographic renormalization scheme. Thus, in the following examples for the AdS black hole (\ref{ADSBH}), the BTZ black hole with angular momentum (\ref{BH-BTZ}), and the charge AdS black hole (\ref{BH-EHW}), we will relate ``$complexity=action$'' where the formalism to obtain at $ {\cal\,T}{\cal\,S}_{W}$ will be more intuitive and explicit to satisfy the first law of thermodynamics through the Wald formalism and holographic renormalization scheme. 

\section{Numerical Applications and Holographic BCFTs}\label{NU}

In this section we apply the Entanglement Entropy Functional (\ref{EEG2}) to the BTZ black hole solution \cite{Iyer:1994ys,Barnich:2007bf,Khodabakhshi:2020fhb,Caceres:2017lbr,Dong:2013qoa,Minamitsuji:2013vra}.  The functional itself has contributions of the AdS/BCFT Horndeski setup. We numerically analyze the dependence on the coupling constant \(\gamma\). Additionally, we discuss the boundary entropy in the context of holographic duals of boundary conformal field theories (BCFTs) through $\Delta_{{\partial\cal\,Q}}$, particularly focusing on the physical implications for Einstein gravity (\( \gamma = 0 \)) and Horndeski gravity with \( \gamma = 0.5 \) and \( \gamma = -0.5 \).

\subsection{3D Planar BTZ Black Hole and Entanglement Entropy}

\begin{figure}[ht]
\centering
\includegraphics[scale=0.5]{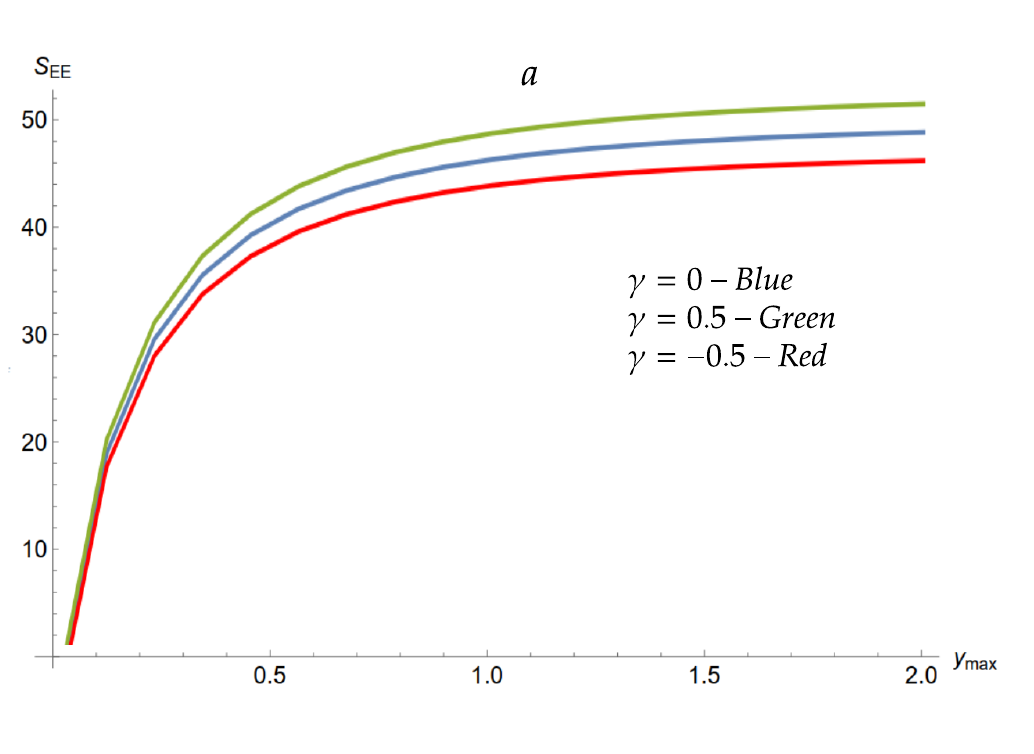}
\includegraphics[scale=0.5]{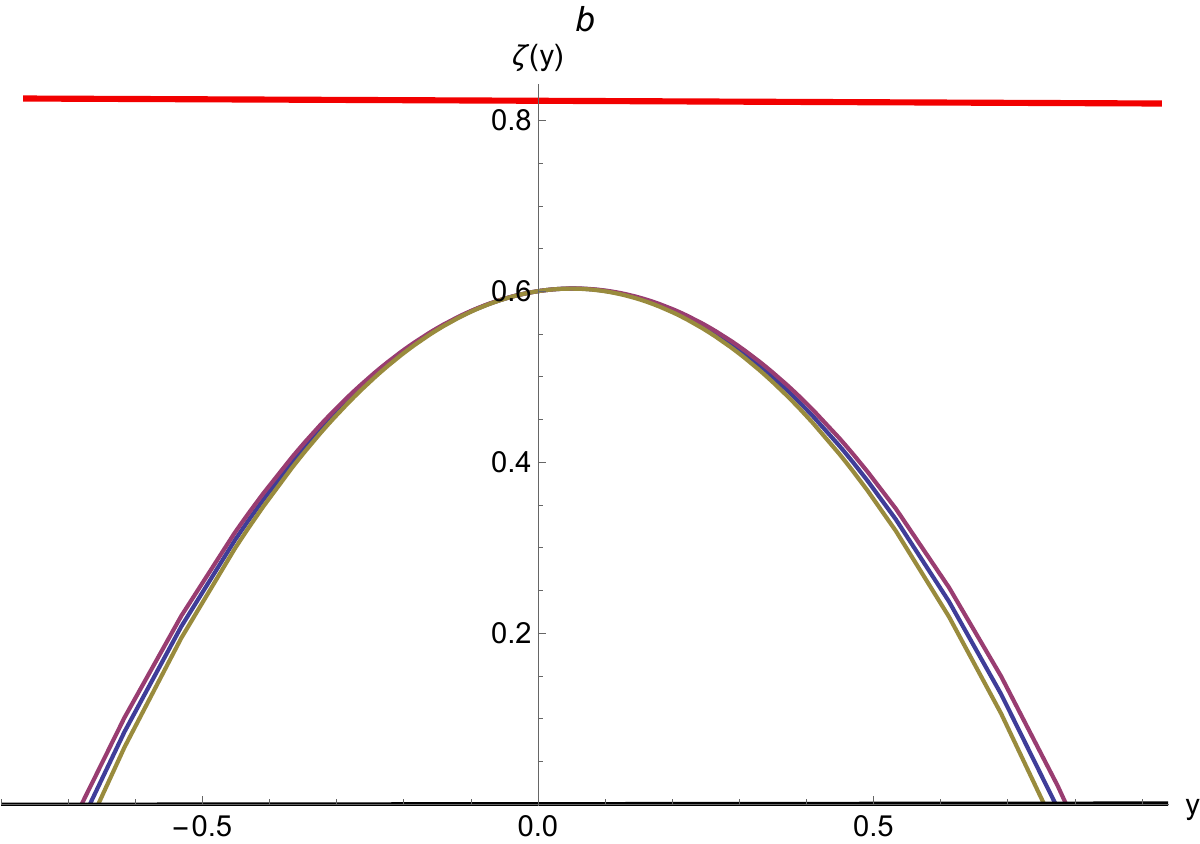}
\caption{Planar black hole in \( d=3 \): (a) Entanglement entropy plotted as a function of the half-width of the boundary interval. (b) Examples of representative minimal surfaces. Both plots are generated using \( g = \mu = \beta = G = 1 \), with the following values of \( \gamma \): \( \gamma = 0 \) (blue), \( \gamma = -0.5 \) (dark green), and \( \gamma =0.5 \) (dark purple).}
\label{fig:streamRT}
\end{figure}

We start considering the 3-dimensional planar black hole (\( d=3 \), \( \epsilon=0 \)), where the metric and scalar fields simplify to:

\begin{eqnarray}
&&ds^2 = -(g^2r^2 - \mu)dt^2 + \frac{dr^2}{g^2r^2 - \mu} + r^2dx^2,\\
&&\phi = \frac{\sqrt{\beta}}{g} \log{\left(gr + \sqrt{(gr)^2 - \mu}\right)} + \phi_0.
\end{eqnarray}
This metric corresponds to the BTZ black hole, with the event horizon located at \( r_+ = \sqrt{\mu}/g \). To easily compute the entanglement entropy, we switch to Fefferman-Graham coordinates \cite{Caceres:2017lbr}, defined as:

\begin{eqnarray}
\zeta = \frac{g}{gr + \sqrt{g^2r^2 - \mu}},    
\end{eqnarray}
and rescale the boundary coordinates as \( \tau = t/2 \) and \( y = x/(2g) \). In these coordinates, the metric becomes:
\begin{eqnarray}
ds^2 = \frac{1}{g^2 \zeta^2} \left[ -(g^2 - \mu \zeta^2)^2 d\tau^2 + d\zeta^2 + (g^2 + \mu \zeta^2)^2 dy^2 \right],    
\end{eqnarray}
while the scalar field satisfies:
\begin{eqnarray}
\frac{d\phi}{d\zeta} = -\frac{\sqrt{\beta}}{g\zeta}.   
\end{eqnarray}
The Ryu-Takayanagi (RT) surface \cite{Santos:2024cwf,Takayanagi:2011zk,Braccia:2019xxi} is parameterized as \( X^\mu = (\tau = \text{const}, y, \zeta(y)) \), and the Entanglement Entropy (${\cal\,S}_{EE}$) functional to be minimized is:
\begin{eqnarray}
&&{\cal\,S}_{EE}=\Delta_{{\cal\,Q}} \int dy \frac{\sqrt{(\zeta')^2 + (g^2 + \mu \zeta^2)^2}}{g\zeta} \left( 1 + \tilde{\gamma} \frac{(\zeta')^2}{(\zeta')^2 + (g^2 + \mu \zeta^2)^2} \right)+\nonumber\\
&&\Delta_{{\partial\cal\,Q}} \int dy \frac{\sqrt{(\zeta')^2 + (g^2 + \mu \zeta^2)^2}}{g\zeta} \left( 1 + \tilde{\gamma} \frac{(\zeta')^2}{(\zeta')^2 + (g^2 + \mu \zeta^2)^2} \right),\label{ETP}   
\end{eqnarray}
where \( \tilde{\gamma} = \frac{\gamma G \beta}{4} \). We numerically minimize this functional and plot the entanglement entropy as a function of the half-width \( y_{\text{max}} \) of the boundary interval (ranging from \( -y_{\text{max}} \) to \( y_{\text{max}} \)) in  Fig. \ref{fig:streamRT}-(a). For comparison of our results, see \cite{Caceres:2017lbr}, we consider three cases: Einstein gravity (\( \gamma = 0 \)), Horndeski gravity with \( \gamma > 0 \), and Horndeski gravity with \( \gamma < 0 \). Although bulk causality typically requires \( \gamma \leq 0 \) \cite{Minamitsuji:2013vra}, we include \( \gamma > 0 \) for completeness. Representative RT surfaces for different values of \( \gamma \) are shown in Fig. \ref{fig:streamRT}-(b). 

\subsection{Boundary Entropy and Physical Implications}
The boundary entropy, a key feature of BCFTs, quantifies the degrees of freedom localized at the boundary. In the holographic dual framework, it's related to the on-shell action of the bulk spacetime with a boundary. For Einstein gravity (\( \gamma = 0 \)), the boundary entropy is consistent with the standard results of holographic BCFTs, as discussed in \cite{Takayanagi:2011zk}. The concavity of the entanglement entropy curves in Fig. \ref{fig:streamRT}-(a) reflects the strong sub-additivity property, a hallmark of entropy in any physical system. This concavity arises naturally from the minimization of the functional entropy of holographic entanglement \cite{Callan:2012ip}.

For Horndeski gravity, the boundary entropy depends on the coupling \( \gamma \). When \( \gamma = 0.5 \), the entropy ${\cal\,S}_{EE}$-(\ref{ETP}) increases, indicating an enhancement of boundary-$\Delta_{{\partial\cal\,Q}}$ degrees of freedom. On the other hand, for \( \gamma = -0.5 \), ${\cal\,S}_{EE}$-(\ref{ETP}) decreases, suggesting suppression of boundary contributions. These results highlight the sensitivity of boundary physics to the choice of bulk-$\Delta_{{\cal\,Q}}$ gravitational theory and its parameters.

\subsection{3D Spherical BTZ Black Hole and Phase Transition}

\begin{figure}[ht]
\centering
\includegraphics[scale=0.35]{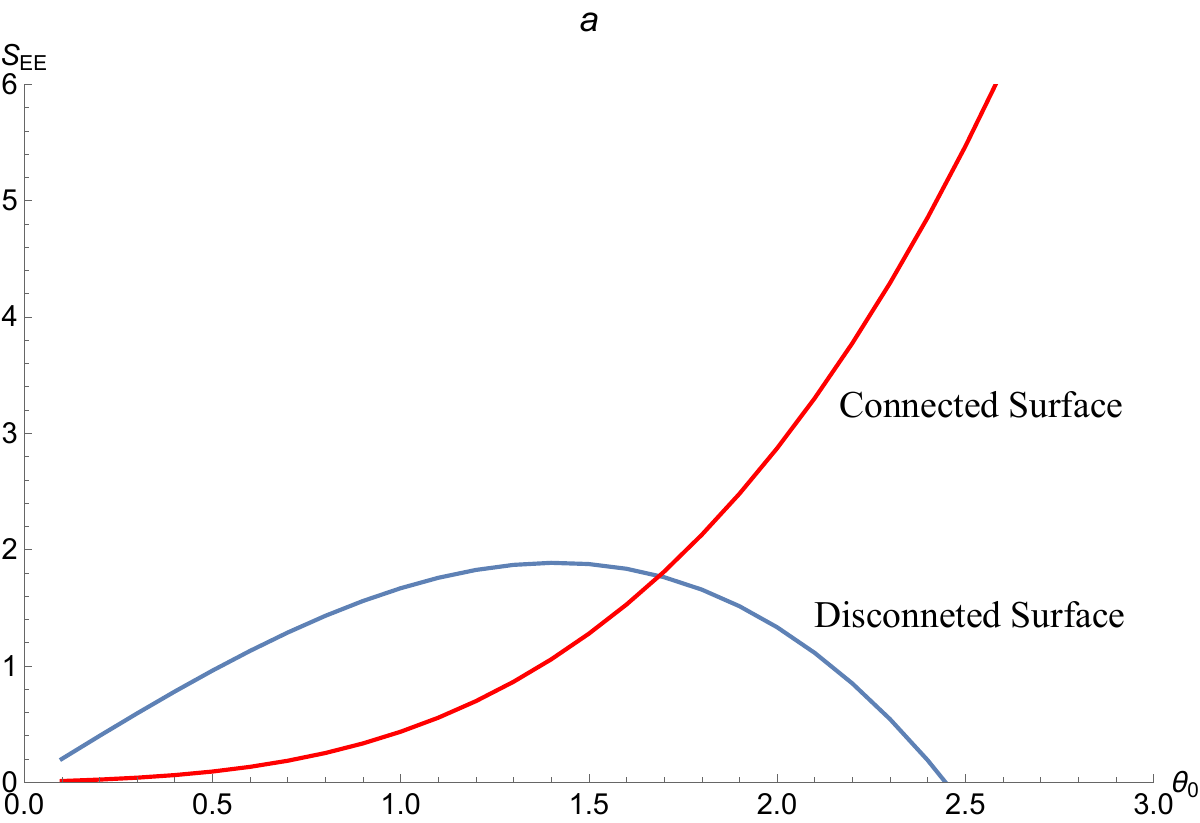}
\includegraphics[scale=0.38]{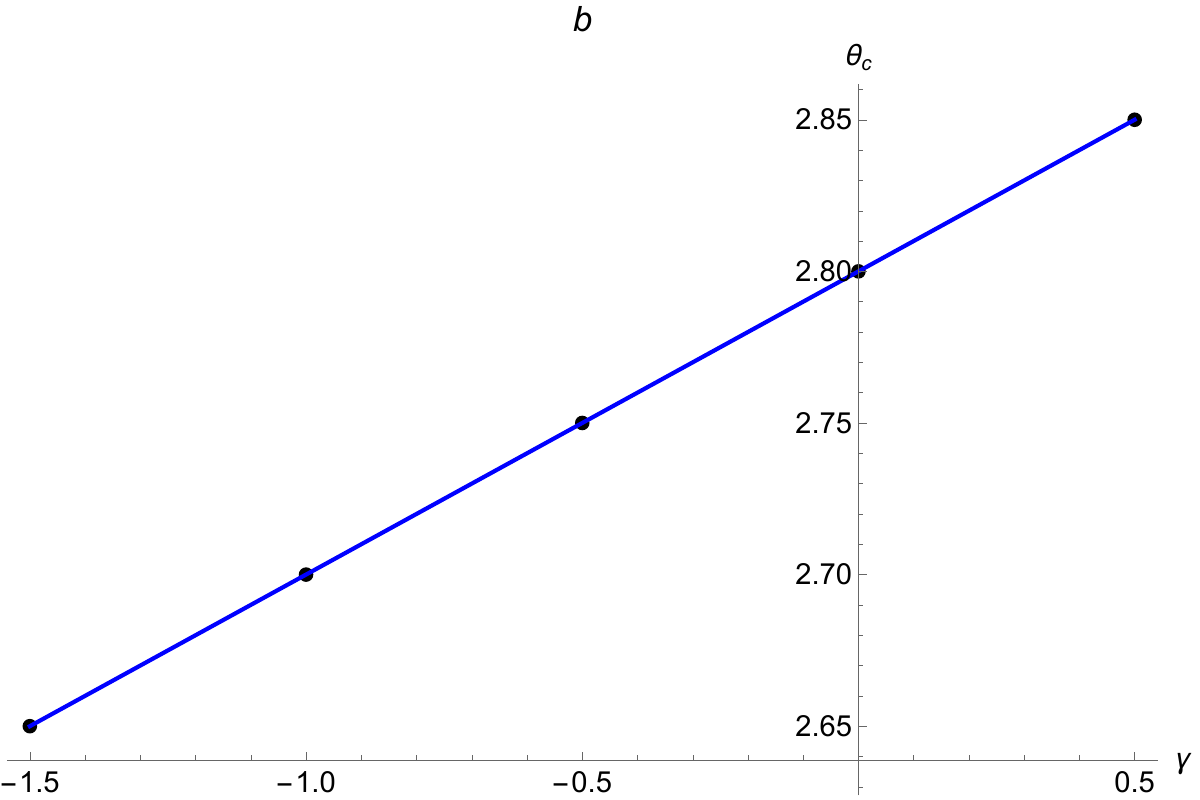}
\caption{For \(d=3\), panel (a) illustrates the Wald entropy as a function of the boundary interval half-width \(\theta_0\). The blue curve represents the connected surface, while the red curve corresponds to the disconnected surface. The parameters used are \(\mu' = g = \beta = G = 1\) and \(\gamma = -0.1\). Panel (b) depicts the relationship between the critical angle \(\theta_c\) and the coupling parameter \(\gamma\), with the same parameter values (\(g = \mu' = \beta = G = 1\)) held constant.}
\label{fig:streamRT1}
\end{figure}

We now turn our attention to the 3-dimensional spherical black hole (\( d=3, \epsilon=1 \)), which retains the BTZ metric structure:
\begin{eqnarray}
ds^2 = -\left(-\mu^\prime + g^2r^2\right)dt^2 + \frac{dr^2}{\left(-\mu^\prime + g^2r^2\right)} + r^2d\varphi^2,    
\end{eqnarray}
where the parameter \(\mu^\prime\) is given by:
\begin{eqnarray}
\mu^\prime = \mu - \frac{16\kappa^2}{(\beta\gamma + 4\kappa)^2}.    
\end{eqnarray}
The scalar field profile remains identical to that of the 3-dimensional planar black hole \cite{Caceres:2017lbr}. For a boundary interval of half-width \(\theta_0\), there are two minimal surfaces satisfying the homology constraint:
\begin{itemize}
\item A connected surface, which spans the boundary interval Fig. \ref{fig:streamRT1}-(a).
\item A disconnected surface, which includes the black hole horizon as a connected component Fig. \ref{fig:streamRT1}-(a).
\end{itemize}
Drawing from intuition in Einstein gravity, a phase transition is expected between these two surfaces as the size of the boundary region (\(\theta_0\)) increases. 

In Fig. \ref{fig:streamRT1}-(a), the size of the boundary interval \(\theta_0\) is varied from \(0\) to \(\pi\), and the functional value is evaluated for the connected and disconnected surfaces. The results reveal:
\begin{itemize}
\item The disconnected surface entropy (\ref{ETP}) increases monotonically with \(\theta_0\), and after \(\theta_c \approx 1.694\) the entropy begins to decline Fig. \ref{fig:streamRT1}-(a), eventually reaching zero when the black hole has fully evaporated all the information is released \cite{Santos:2024cvx}. 
\item The connected surface entropy (\ref{ETP}) increases as \(\theta_0\) increases.
\end{itemize}
The two curves intersect at a critical angle \(\theta_c \approx 1.694\) radians, marking the phase transition. Beyond this angle, the disconnected surface produces a smaller value of the functional, and the homology constraint dictates that it becomes the dominant contribution \cite{Caceres:2017lbr,Minamitsuji:2013vra,Callan:2012ip}. This phase transition has significant implications for the Araki-Lieb inequality, which states:
\begin{eqnarray}
|{\cal\,S}_{{\cal\,A}}-{\cal\,S}^{C}_{{\cal\,A}}| \leq {\cal\,S}^{BH}_{\mathrm{thermal}},    
\end{eqnarray}
where \({\cal\,S}_{{\cal\,A}}\) and \({\cal\,S}^{C}_{{\cal\,A}}\) are the entropies of a subregion \({\cal\,A}\) and its complement, respectively, and \({\cal\,S}^{BH}_{\mathrm{thermal}}\) is the thermal entropy. For \(\theta_0 \geq \theta_c\), the phase transition ensures that the difference between \({\cal\,S}_{{\cal\,A}}\) and \({\cal\,S}^{C}_{{\cal\,A}}\) equals the thermal entropy, saturating the inequality. This saturation manifests itself as an entanglement plateau when plotting \(|{\cal\,S}_{{\cal\,A}}-{\cal\,S}^{C}_{{\cal\,A}}|/{\cal\,S}^{BH}_{\mathrm{thermal}}\) against \(\theta_0\) \cite{Caceres:2017lbr}.

To further explore the role of the coupling parameter \(\gamma\), we fix \(g\), \(\mu\), and \(\beta\) to unity and vary \(\gamma\). For each value of \(\gamma\), the critical angle \(\theta_c\) is computed and plotted in Fig. \ref{fig:streamRT1}-(b). The results focus on the negative regime \(\gamma\), consistent with the bulk causality constraints \cite{Caceres:2017lbr,Minamitsuji:2013vra}. As \(\gamma\) becomes more negative, the critical angle \(\theta_c\) increases, leading to a larger entanglement plateau \cite{Santos:2024cvx}. For \(\gamma = 0\), the results align with the analytical value of \(\theta_c\) derived for the BTZ black hole in \cite{Hubeny:2013gta}.

\subsection{Relating Entanglement Entropy, Phase Transitions, and Complexity Growth}

The study of entanglement entropy in 3-dimensional BTZ black holes, particularly in the context of Horndeski gravity, provides a deeper understanding of the interplay between boundary physics and bulk gravitational dynamics \cite{Santos:2024cvx}. This framework naturally connects to the conjecture of linear complexity growth, as proposed in Nielsen's geometrization of complexity framework \cite{Nielsen:2005mkt}, and its thermodynamic implications.

The entanglement entropy functional, as derived for both planar and spherical black holes, reflects the intricate relationship between the geometry of the bulk spacetime and the degrees of freedom on the boundary. In Horndeski gravity, the coupling parameter \(\gamma\) introduces modifications to the entanglement entropy, which, in turn, affects the boundary entropy and the critical angle \(\theta_c\) of phase transitions \cite{Caceres:2017lbr,Minamitsuji:2013vra}. These modifications are not only geometric but also encode information about the underlying thermodynamic properties of the system.

In the context of complexity growth, the linear growth conjecture posits that the complexity of a quantum system increases linearly with time until it saturates at a maximum value \cite{Susskind:2014rva,Brown:2015bva,Brown:2015lvg}. This growth is hypothesized to be proportional to the product of black hole entropy \(\mathcal{S}_{BH}\) and temperature \({\cal T}\), providing a natural scale for near-extremal black holes. The functional entanglement entropy in Horndeski gravity, which depends on \(\gamma\), \(\mathcal{S}_{BH}\), and \({\cal T}\), aligns with this conjecture by offering a holographic perspective on the growth of complexity.

The phase transition between connected and disconnected minimal surfaces, as observed in the spherical BTZ black hole case, provides a geometric manifestation of the transition from linear complexity growth to saturation. For \(\theta_0 < \theta_c\) \cite{Caceres:2017lbr}, the connected surface dominates, corresponding to a regime where the complexity grows linearly \cite{complexityshocks,Susskind:2014moa,Susskind:2014rva,Brown:2015bva,Brown:2015lvg}. As \(\theta_0\) approaches \(\theta_c\), the system transitions to a disconnected surface, marking the onset of complexity saturation. This transition is analogous to the saturation of complexity as the system approaches its maximum value, as suggested in \cite{Brown:2015bva}.

Interestingly, extremal black holes, which serve as ground states, do not exhibit growth in complexity. This is consistent with the observation that the functional entanglement entropy for extremal black holes lacks the temperature-dependent term \({\cal T}\), highlighting the absence of thermodynamic driving forces for complexity growth. In this sense, the entanglement plateau observed for \(\theta_0 > \theta_c\) in the spherical black hole case mirrors the behavior of extremal black holes, where complexity remains constant.

\section{The Two-sides Cases}\label{BHSL}

As discussed in Sec. \ref{v0}, in general Horndeski theories the characteristic hypersurfaces that determine the domain of dependence – and thus the WdW patch – need not coincide with the null cones of the physical metric \cite{Kovacs:2020pns,Papallo:2017qvl,Bettoni:2016mij}. In this section we restrict to parameter regimes and symmetric backgrounds for which the fastest propagating modes can be effectively described by null hypersurfaces of an appropriate effective metric. In practice we represent the WdW patch using null rays of the metric (\ref{geomet}) and its BTZ and charged analogues, implicitly assuming that this effective null structure captures the relevant causal domain for the modes that govern holographic complexity. 

\begin{figure}[ht]
\centering
\includegraphics[scale=0.95]{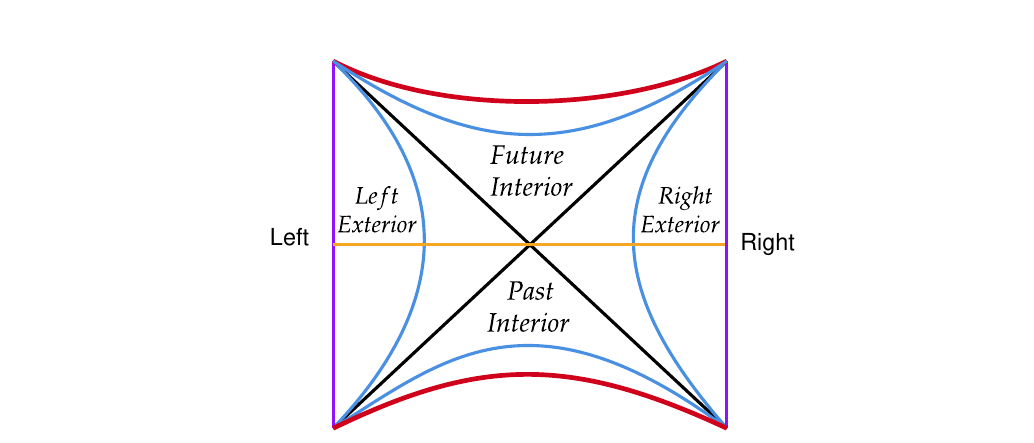}
\caption{The Penrose diagram for an AdS two-sided black hole with $t=\tau$ (right side) and $t=-\tau$ (left side).}
\label{fig:stream4}
\end{figure}

In the two-sided AdS black hole, the geometry features two asymptotic regions, each containing an identical black hole \cite{Hartman:2013qma,Susskind:2014moa}. These regions, referred to as the left and right sides in Fig. \ref{fig:stream4}, are connected by an ERB \cite{Susskind:2014yaa}. The nontrivial topology of the ERB is present from the outset, eliminating the need for a topological transition. 

The AdS/BCFT correspondence \cite{Santos:2024cwf,Takayanagi:2011zk,Santos:2024cvx} provides a natural framework for studying ERB \cite{Susskind:2014yaa} with multiple asymptotic boundaries. In this context, the bulk geometry (AdS spacetime) encodes information about the Boundary Conformal Field Theory (BCFT). The ERB connecting two asymptotic regions corresponds to the entanglement between the two boundary theories \cite{Santos:2024cvx}. This connection is particularly relevant in Horndeski gravity, where including scalar-tensor interactions modifies the bulk dynamics and consequently, the boundary entanglement structure. By analyzing ERBs in this extended framework, one can probe the effects of modified gravity on holographic entanglement entropy and other boundary observables.

\subsection{AdS Black Hole}\label{ADSBH} 
The metric describing the geometry of Fig. \ref{fig:stream4} has the form,
\begin{eqnarray}
dS^2_{\mathcal{Q}}=\frac{L^2}{r^2}\left(-f(r)dt^2+dx^2+dy^2+\frac{dr^2}{f(r)}\right),\label{geomet}
\end{eqnarray}
where the induced metric for $dS^2_{\mathcal{Q}}$ is 
\begin{eqnarray}
dS^2_{\partial\mathcal{Q}}=\frac{L^2}{r^2}\left(-f(r)dt^2+dx^2+\frac{g^2(r)dr^2}{f(r)}\right),
\end{eqnarray}
with $g^2(r)=1+y{'}(r)f(r)$ and $y{'}(r)=dr/dy$. The equation of motion for the scalar field (${\cal E}^{\mathcal{Q}}_{\phi}=0$) admits a first integral (${\cal E}^{\mathcal{Q}}_{\phi}={\cal J}_0$), which implies ${\cal J}^{\mu\nu}={\cal J}_0$ with ${\cal J}_0$ being a constant. Then, the solution for $dS^2_{\mathcal{Q}}$ with ${\cal J}_0=0$ and ${\cal E}^{\mathcal{Q}}_{\mu\nu}=0$ is
\begin{eqnarray}
&&f(r)= 1-\left(\frac{r}{r_{h}}\right)^{3},\label{Solbulk}\\
&&\psi^{2}(r)=(\phi'(r))^2=-\frac{2L^{2}\xi}{\gamma r^{2}f(r)}\,\,;\,\,\xi=\frac{1
}{2}\frac{\alpha+\gamma\Lambda}{\alpha},
\end{eqnarray}
In the above equation, we apply the rescaling of \cite{Santos:2024zoh,Santos:2021orr,Santos:2023flb}, and for $dS^2_{\partial\mathcal{Q}}$ with ${\cal F}^{\partial\mathcal{Q}}_{\phi}={\cal G}_{0}$, which provides ${\cal G}_{\mu\nu}={\cal G}_{0}$ where ${\cal G}_0$ is a constant. Here, for simplicity, we choose ${\cal G}_0=0$ in equation (\ref{L10}), we have the solution

\begin{eqnarray}
y(r)=y_0+\int{\frac{(\Sigma\,L)dr}{\sqrt{4-(\Sigma\,L)^2f(r)}}}.\label{Solbound}
\end{eqnarray}
Here, conditions ${\cal E}^{\mathcal{Q}}_{\phi}=0$ and ${\cal F}^{\partial\mathcal{Q}}_{\phi}=0$ align themselves with the "no-hair theorems" that assert that the properties of the black hole - defined as a black hole with a vacuum exterior and interior - are fully characterized by three parameters: the mass \( {\cal\,M} \), angular momentum \( J \), and electric charge \( Q \) \cite{Brown:2015lvg}. The choice of ${\cal J}_0=0$ and ${\cal G}_0=0$ provides consistency for the product of entropy and temperature; for more details, see \cite{Susskind:2014rva,Nielsen:2005mkt}. For the cases (\ref{BH-BTZ}), (\ref{BH-EHW}), and (\ref{TSHO}), we will execute the same procedure. Now, combining (\ref{Solbulk}) and (\ref{Solbound}) in (\ref{1}), we have
\begin{eqnarray}
&&\dot{{\cal A}}_{\mathcal{Q}}(t_L+t_R)=\frac{d{\cal A}_{\mathcal{Q}}}{d(t_L+t_R)}=\frac{1}{\,r_h}\left[\frac{\Delta_{{\cal\,Q}}}{4r^2_hG_{N}}\left(1+\frac{\xi}{4}\right)\right],\label{CP1}\\
&&\dot{{\cal A}}_{\partial\mathcal{Q}}(t_L+t_R)=\frac{d{\cal A}_{\partial\mathcal{Q}}}{d(t_L+t_R)}=\frac{1}{\,r_h}\left[\frac{\Delta_{{\cal\partial\,Q}}}{r^2_hG_{N}}\left(1+\frac{\xi}{4}\right)\right],\label{CP2}
\end{eqnarray}
where the black hole entropy ($\mathcal{S}_{BH}=\mathcal{S}^{\mathcal{Q}}_{BH}+\mathcal{S}^{\partial\mathcal{Q}}_{BH}$-\cite{Santos:2024zoh,Santos:2021orr,Santos:2023flb,Hajian:2020dcq,Feng:2015oea}) is obtained through the derivation of entropy in Horndeski presented in Sec. \ref{WEHT}:
\begin{eqnarray}
&&\mathcal{S}^{\mathcal{Q}}_{BH}=\frac{\Delta_{{\cal\,Q}}}{4r^2_hG_{N}}\left(1+\frac{\xi}{4}\right),\\
&&\mathcal{S}^{\partial\mathcal{Q}}_{BH}=\frac{\Delta_{{\cal\partial\,Q}}}{r^2_hG_{N}}\left(1+\frac{\xi}{4}\right).
\end{eqnarray}
The temperature is given by ${\cal T}\to\frac{1}{\,r_h}$. So, plugging $\mathcal{S}_{BH}=\mathcal{S}^{\mathcal{Q}}_{BH}+\mathcal{S}^{\partial\mathcal{Q}}_{BH}$ into the equations (\ref{CP1})-(\ref{CP2}), we have
\begin{eqnarray}
&&\dot{{\cal A}}(t_L+t_R)=\dot{{\cal A}}^{{\cal Q}}(t_L+t_R)+\dot{{\cal A}}^{{\cal\partial\,Q}}(t_L+t_R)={\cal T}{\cal S}_{BH}\\
&&\,\,\,\,\,\,\,\,\,\,\,\,\,\,\,\,\,\,\,\,\,\,\,\,\,\,\,\,\,\,\,\,\,\,\to{\cal C}(t)=\frac{2{\cal T}\,\mathcal{S}_{BH}}{\pi\hbar}
\end{eqnarray}
The conjecture that complexity grows linearly for a significant period, as proposed in Nielsen's geometrization of complexity framework \cite{Nielsen:2005mkt}, has been supported by various studies. However, the precise duration of this linear growth remains an open question. It's hypothesized that growth persists until complexity approaches its maximum value, as suggested in \cite{Susskind:2014rva}. Here, our derivation of the linear growth is proportional to the entropy $\mathcal{S}_{BH}$ in the Horndeski gravity framework as a product of entropy and temperature (${\cal\,T}\,\mathcal{S}_{BH}$), providing a natural scale for near-extremal black holes. 

Using the Wald prescription presented in Sec. \ref{WEHT}, it is easy to verify that the first law is also satisfied as ${\cal T}{\cal S}_{BH}={\cal M}$ where ${\cal M}$ is the black hole mass. In this framework, the geometry of an Einstein-Rosen bridge can be associated with maximal spatial volume slices Fig. \ref{CPX}, while the rate of complexity growth is conjectured to be bounded by the product of entropy and temperature: 
\begin{eqnarray}
\mathcal{C}(t_R+t_L)=\frac{2{\cal T}\,\mathcal{S}_{BH}}{\pi\hbar}.
\end{eqnarray}
 For black holes, this bound is saturated, providing a natural connection between thermodynamic quantities and holographic complexity \cite{Brown:2015lvg,Roberts:2014ifa,Shenker:2013pqa,Susskind:2014rva}.

The spacetime volume \(|\mathcal{W}|\) of the Wheeler-DeWitt patch and the spatial volume \({\cal V}\) of the maximal slice are related by \(|\mathcal{W}| \sim {\cal V} L\) Fig. \ref{CPX}, where \(L\) is the AdS curvature scale \cite{Brown:2015lvg}. This relationship holds for large AdS-Schwarzschild black holes, with corrections arising from the Horndeski parameters \(\alpha\) and \(\gamma\), which modify the effective geometry and thermodynamic properties of the black hole.

The product \({\cal TS}_{BH}\) remains a robust measure of complexity growth, even in the presence of Horndeski corrections. For neutral black holes, the rate of change of action is proportional to the black hole mass, with the product \({\cal TS}_{BH}\)  and the mass, that is, \({\cal TS}_{BH}={\cal M}\). The rate of change of action remains within a factor of a few of \({\cal TS}_{BH}\):
\begin{itemize}
\item \(E_\psi={\cal TS}_{BH}=\frac{4}{3}{\cal M}\) \(\left(\to\mathcal{C}(t)=\frac{8}{3\pi\hbar}{\cal M}\right)\) (large black hole , \(r_h\gg\,L\)),
\item \(E_\psi={\cal TS}_{BH}=\frac{2}{3}{\cal M}\) \(\left(\to\mathcal{C}(t)=\frac{4}{3\pi\hbar}{\cal M}\right)\) (small black hole , \(r_h\ll\,L\)),
\end{itemize}
even when Horndeski corrections are included \cite{Lloyd:2000cry}.

The \({\cal CA}\)-duality provides a degree of universality absent in the \({\cal CV}\)-duality. In particular, the universal constant connecting action and complexity remain consistent across all neutral black holes studied, including those with Horndeski modifications. This universality underscores the robustness of the ${\cal CA}$-duality framework in capturing the holographic complexity of black holes, even in extended theories of gravity.

\subsection{BTZ black hole}\label{BH-BTZ}
\begin{figure}[!ht]
\begin{center}
\includegraphics[scale=0.85]{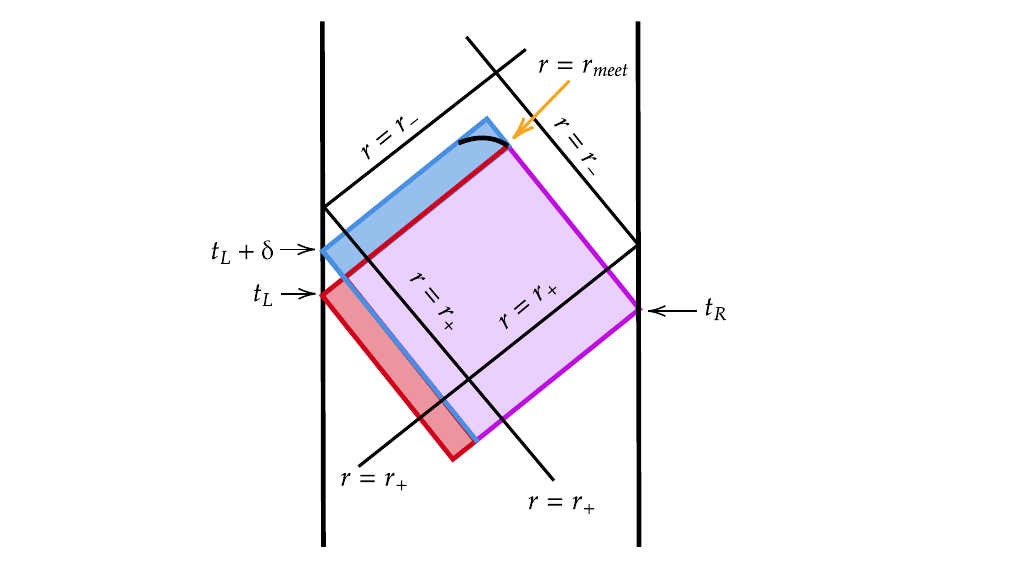}
\caption{The Wheeler-DeWitt patch, a crucial construct for understanding the causal structure of the spacetime of the BTZ black hole, evolves dynamically as the left boundary time \(t_L\) increases. Specifically, the patch gains a segment (depicted in blue) and simultaneously loses another (depicted in red), with the transition occurring at the radial coordinate \(r = r_\textrm{meet}(t_L, t_R)\).}
\label{pensr}
\end{center}
\end{figure}

For rotating black holes in 2+1-dimensional AdS (BTZ black hole), we have

\begin{eqnarray}
dS^2_{\mathcal{Q}}=-f(r)dt^{2}+r^{2}\left(dy-\frac{J}{r^{2}}dt\right)^{2}+\frac{dr^{2}}{f(r)}
\end{eqnarray}
where the induced metric for $dS^2_{\partial\mathcal{Q}}$ is 
\begin{eqnarray}
dS^2_{\partial\mathcal{Q}}=-\left(f(r)-\frac{J^2}{r^2}\right)dt^{2}+\frac{g^{2}(r)dr^{2}}{f(r)}-2Jdydt
\end{eqnarray}
with $g^2(r)=1+r^2y{'}(r)f(r)$ and $y{'}(r)=dr/dy$. Then, the solution for $dS^2_{\mathcal{Q}}$ with ${\cal E}^{\mathcal{Q}}_{\phi}=0={\cal E}^{\mathcal{Q}}_{rr}$ is
\begin{eqnarray}
&&f(r)=-m^2+\frac{\alpha r^{2}}{\gamma}+\frac{j^{2}}{r^{2}},\label{Solbulk1}\\
&&\psi^{2}(r)=(\phi'(r))^2=-\frac{2\xi}{\gamma\,f(r)}\,\,;\,\,\xi=\frac{1
}{2}\frac{\alpha+\gamma\Lambda}{\alpha},
\end{eqnarray}
where we can write the mass, the angular momentum, the angular velocity of the horizon, the surface gravity, and the radii of the horizon as \cite{Hajian:2020dcq,Feng:2015oea}:
\begin{align}
&{\cal M}=\frac{(\alpha -\Lambda \gamma)m}{16 \alpha\,G_N}, \qquad J=   \frac{(\alpha -\Lambda \gamma)j}{8 \alpha\,G_N},\nonumber\\
& \kappa_\pm=\frac{\alpha(r^2_+-r^2_-)}{\gamma r_\pm}, \qquad \Omega_\pm=\frac{j}{r_\pm^2},  \nonumber\\
& r_\pm^2=\frac{\gamma m \mp \sqrt{\gamma^2m^2-4\gamma\alpha j^2}}{2\alpha}. 
\end{align}
For $dS^2_{\partial\mathcal{Q}}$, we have

\begin{eqnarray}
y{'}(r)=\frac{(\Sigma\,l_{AdS})}{\sqrt{4-(\Sigma\,l_{AdS})^2f(r)}},\label{Solbound1}
\end{eqnarray}
combining (\ref{Solbulk1}) and (\ref{Solbound1}) in (\ref{1}) for the BTZ black hole, we have
\begin{eqnarray}
&&\dot{{\cal A}}_{\mathcal{Q}}(t_L+t_R)=\frac{d{\cal A}_{\mathcal{Q}}}{d(t_L+t_R)}=\frac{\Delta_{{\cal\,Q}}\,(r^2_{+}-r^{2}_{-})}{4G_{N}},\\
&&\dot{{\cal A}}_{\partial\mathcal{Q}}(t_L+t_R)=\frac{d{\cal A}_{\partial\mathcal{Q}}}{d(t_L+t_R)}=\frac{2\Delta_{{\cal\partial\,Q}}\,(r^2_{+}-r^{2}_{-})}{G_{N}}.
\end{eqnarray}
Here, we can use
\begin{eqnarray}
&&\frac{\Delta_{{\cal\,Q}}(r^2_{+}-r^{2}_{-})}{4G_{N}}=({\cal M}-\Omega\,J)_{{\cal Q}}={\cal T}{\cal S}^{{\cal Q}}_{BH},\\
&&\frac{2\Delta_{{\cal\partial\,Q}}(r^2_{+}-r^{2}_{-})}{G_{N}}=({\cal M}-\Omega\,J)_{\partial{\cal Q}}={\cal T}{\cal S}^{{\cal \partial\,Q}}_{BH}.
\end{eqnarray}
The first law of thermodynamics is satisfied, i.e. ${\cal M}-\Omega\,J={\cal T}{\cal S}_{BH}$ \cite{Hajian:2020dcq,Feng:2015oea} with ${\cal S}_{BH}={\cal S}^{{\cal Q}}_{BH}+{\cal S}^{{\cal \partial\,Q}}_{BH}$; we apply the Wald formulation present in Sec. \ref{WEHT}. The conjecture complexity is given by
\begin{eqnarray}
{\cal C}(t)=\frac{2}{\pi\hbar}\left[\frac{d{\cal A}_{\mathcal{Q}}}{d(t_L+t_R)}+\frac{d{\cal A}_{\partial\mathcal{Q}}}{d(t_L+t_R)}\right]=\frac{2{\cal T}\,\mathcal{S}_{BH}}{\pi\hbar},
\end{eqnarray}
where the addition of angular momentum to an AdS black hole modifies the geometry of the Wheeler-DeWitt (WdW) patch as shown in Fig. \ref{pensr}. The WdW patch, instead of singularity at \( r=0 \), ends when the incoming light sheets collide at \( t=0 \) (for \( t_L = t_R \)) \cite{Brown:2015lvg}. This has profound implications for the late-time growth of complexity, which can be analyzed using the framework of the complexity conjecture \cite{Susskind:2014rva} and the methodology outlined in \cite{Santos:2020xox}.

\begin{figure}[!ht]
\begin{center}
\includegraphics[scale=0.85]{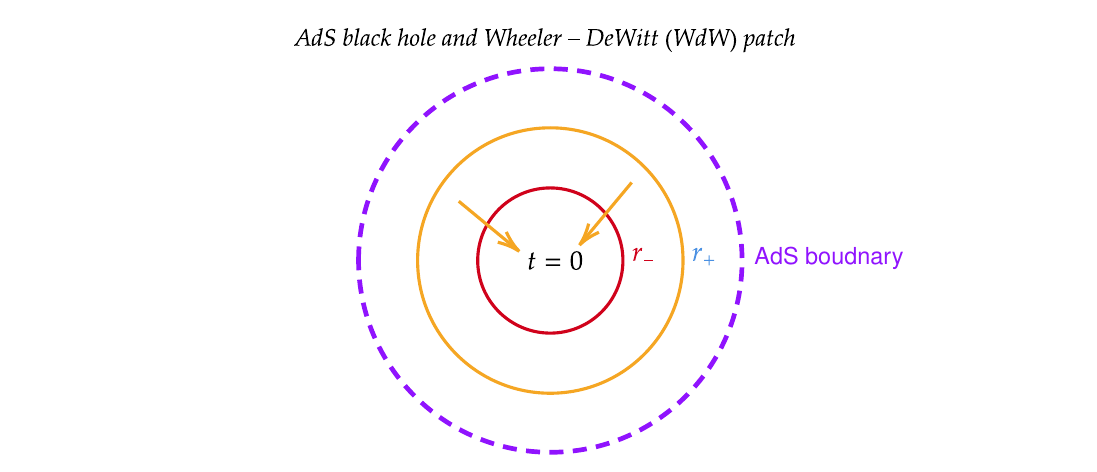}
\caption{The diagram has been constructed and visually represents the AdS black hole geometry, including the outer and inner horizons (${\color{blue}r_{+}}$  and ${\color{red}r_{-}}$), the Wheeler-DeWitt (WdW) patch, and the collision of light sheets at $t=0$.}
\label{WB}
\end{center}
\end{figure}

Furthermore, within the framework of Horndeski gravity, the relationship between angular momentum and AdS boundary conditions introduces additional corrections to the black hole's causal structure and the holographic complexity \cite{Santos:2024zoh,Santos:2020xox}. These corrections can be interpreted as boundary deformations in the AdS/BCFT correspondence, where the boundary conformal field theory encodes the effects of angular momentum and second-order derivative terms in the bulk action. The late-time behavior of complexity, governed by the volume of the WdW patch Fig. \ref{WB}, provides a direct probe of these deformations and their impact on the dual field theory.

\subsection{Einstein-Horndeski-Maxwell action}\label{BH-EHW}

Now, we consider a system whose low-energy bulk physics is described by the Einstein-Horndeski-Maxwell action:

\begin{eqnarray}
&&\mathcal{A}=\frac{1}{\kappa}\int_{\mathcal{Q}}\sqrt{|g|}\; {\cal L}_{\rm H}+\int_{\mathcal{Q}}\sqrt{|g|}\; {\cal L}_{\rm {\cal M}}-\frac{1}{4}\int_{\mathcal{Q}}\sqrt{|g|}{F^2}\cr 
&&+\frac{2}{\kappa}\int_{\partial\mathcal{Q}}{\sqrt{|h|}\mathcal{L}_{bdry}}+\int_{\partial\mathcal{Q}}{\sqrt{|h|}\mathcal{L}_{mat}}+\frac{2}{\kappa}\int_{ct}{\sqrt{|h|}\mathcal{L}_{ct}}.\label{EHW}
\end{eqnarray}
The objective of the study of (\ref{EHW}), is to extend the findings presented \cite{Brown:2015bva} by including the electric field in the framework of Horndeski gravity \cite{Hajian:2020dcq,Feng:2015wvb}. This extension is motivated by the need to explore the gauge fields and spherically symmetric spacetimes, which have been a topic of significant interest \cite{Melnikov:2012tb,Hartnoll:2009sz}. With the Maxwell term, we aim to provide a more comprehensive framework that captures the dynamics of such systems and contributes to the broader understanding of gauge-gravity interactions.

To analyze the action (\ref{EHW}), we consider a gauge field ${\cal A}={\cal A}_0(r)dt$ \cite{Jain:2015txa,Nashed:2018cth,Nashed:2020kdb}. This electric field, in the context of the conjecture ${\cal C}{\cal A}$, provides a controlled way of studying the relationship between complexity and bulk geometry. For the Horndeski scenario, this approach offers a novel perspective on the role of scalar fields and the dynamics of complexity.

Here, we work with the following metric:
\begin{eqnarray}
dS^2_{\mathcal{Q}}=-h(r)dt^2+r^2(dx^2+dy^2)+\frac{dr^2}{f(r)}.\label{geomet1}
\end{eqnarray}
As we know, the equation of motion for the scalar field ${\cal E}^{\mathcal{Q}}_{\phi}$ and ${\cal F}^{\partial\mathcal{Q}}_{\phi}$ admits a first integral \cite{Hajian:2020dcq,Feng:2015wvb}, which implies in the equations ${\cal E}^{\mathcal{Q}}_{\phi}={\cal J}_0$ and ${\cal F}^{\partial\mathcal{Q}}_{\phi}={\cal G}_0$. To construct reasonable solutions, we consider ${\cal J}_0=0$ and ${\cal G}_0=0$, and then find
\begin{eqnarray}
&&Bulk-f(r)=\frac{\alpha\,r^2h(r)}{\gamma  \left(h(r)+r h'(r)\right)},\\
&&Boundary-y(r)=y_0+\int{\frac{(\Sigma\,L)dr}{\sqrt{4-(\Sigma\,L)^2f(r)}}},
\end{eqnarray}
where the Maxwell equation $\nabla^\mu\,{\cal F}_{\mu\nu}=0$ admits a first integral as
\begin{eqnarray}
{\cal A}^{'}_0(r)=\frac{Q}{r^2}\sqrt{\frac{h(r)}{f(r)}}
\end{eqnarray}
Although the derivative of the scalar field becomes unbounded at the horizon, the quantity \(\gamma\nabla^\mu\phi\nabla_\mu\phi = \frac{-Q^2}{2r^2}\) remains finite at the horizon.

The metric functions for the solution are given by \cite{Hajian:2020dcq,Feng:2015oea}:
\begin{eqnarray}
h = \frac{\alpha r^2}{3\gamma} - \frac{m}{r} + \frac{Q^2}{r^2} - \frac{3\gamma\,Q^4\ell^2}{2\alpha\,r^6}, \quad f = \frac{16r^8 h}{\left(\frac{2\gamma\,Q^2\ell^2}{\alpha} -4r^4\right)^2},  
\end{eqnarray}
with the scalar field and gauge potential expressed as:
\begin{eqnarray}
d\phi = \sqrt{ - \frac{2\gamma\,Q^2\ell^2}{\alpha fr^4}} \, dr, \quad A = \left(\frac{Q}{r} - \frac{\gamma\,Q^3\ell^2}{10\alpha r^5}\right) dt.    
\end{eqnarray}
The brane is located at \(r = r_h\), where the conditions \(f(r_h) = h(r_h) = 0\) are satisfied. The mass and electric charge densities for this configuration are:
\begin{eqnarray}
{\cal M} = \frac{(\alpha+\gamma\Lambda)m}{32\alpha\pi G_N}.    
\end{eqnarray}
The surface gravity and electric potential at the horizon are:
\begin{eqnarray}
\kappa = \frac{\alpha r_h}{2\gamma} - \frac{Q^2}{4r^3_h}, \quad \Phi_{_\text{h}} = \frac{Q}{r_h} - \frac{\gamma\,Q^3}{10\alpha\,r^5_h}.    
\end{eqnarray}
Using the Hawking temperature \({\cal T}_0 = \frac{\kappa}{2\pi}\), the first law of thermodynamics is not satisfied, and the charge associated with the vector \(\frac{1}{T_0}\xi_{_\text{H}}\) is non-integrable \cite{Feng:2015wvb}. However, a modified temperature can be defined as:
\begin{eqnarray}
{\cal T}= \left(\frac{12r^4_h  - \frac{6\gamma\,Q^2}{\alpha}}{12r^4_h}\right) T_0,    
\end{eqnarray}
which ensures the integrability of the entropy. The first law of thermodynamics is then satisfied in terms of the charge densities:
\begin{eqnarray}
{\cal T} \delta{\cal S}_{\text{BH}}={\cal T}( \delta{\cal S}^{{\cal Q}}_{\text{BH}}+\delta{\cal S}^{{\cal\partial Q}}_{\text{BH}} )= \delta{\cal M} - \Phi_h \delta Q.    
\end{eqnarray}
The condition $h(r)=f(r)$, analyze the bulk geometry and its maximal volume slices more tractably \cite{Hartnoll:2009sz}. The calculation of complexity provides clearer insight into the relationship between the field and the bulk geometry. With this calculation of complexity, provided through the holographic renormalization scheme, we have:
\begin{eqnarray}
&&\dot{{\cal A}}_{\mathcal{Q}}(t_L+t_R)=\frac{d{\cal A}_{\mathcal{Q}}}{d(t_L+t_R)}={\cal T}\delta{\cal S}^{{\cal Q}}_{\text{BH}},\\
&&\dot{{\cal A}}_{\partial\mathcal{Q}}(t_L+t_R)=\frac{d{\cal A}_{\partial\mathcal{Q}}}{d(t_L+t_R)}={\cal T}\delta{\cal S}^{{\cal\partial Q}}_{\text{BH}},\\
\end{eqnarray}
No extra counterterms are required since, in dimensions $d\geq\,3$, the Maxwell field diminishes rapidly enough as it approaches the boundary \cite{Hartnoll:2009sz}. So, the entropy is given by:
\begin{eqnarray}
&&\mathcal{S}^{\mathcal{Q}}_{BH}=\frac{r^2_h\Delta_{{\cal\,Q}}}{4G_{N}}\left(1+\frac{\xi}{4}\right)\\
&&\mathcal{S}^{\partial\mathcal{Q}}_{BH}=\frac{r^2_h\Delta_{{\cal\partial\,Q}}}{G_{N}}\left(1+\frac{\xi}{4}\right)
\end{eqnarray}
where $\dot{{\cal A}}(t_L+t_R)=\dot{{\cal A}}^{{\cal Q}}(t_L+t_R)+\dot{{\cal A}}^{{\cal\partial\,Q}}(t_L+t_R)={\cal T}{\cal S}_{BH}$. Now, we can write that
\begin{eqnarray}
{\cal C}(t)=\frac{2{\cal T}\,\mathcal{S}_{BH}}{\pi\hbar}
\end{eqnarray}
As we show, including a Maxwell term in the Einstein-Horndeski action provides a richer framework for exploring the dynamics of gauge-gravity systems. The results establish a direct connection between the ``$complexity=action$'' conjecture and the thermodynamic properties of black holes in Horndeski gravity, particularly in the presence of an electric field \cite{Susskind:2018fmx}. 

\section{Testing our derivation of ${\cal C}{\cal A}$ conjecture}\label{TEST}

For our derivation of the duality of conjecture complexity, ($Complexity=Action$) ${\cal C}{\cal A}$ duality, we define that the growth of complexity is given by
\begin{eqnarray}
\mathcal{C}(e^{-iHt}\ket{\psi})\leq\frac{2E_{\psi}}{\pi\hbar}\,;\,\,{\cal C}_\psi(t)=\frac{d{\cal C}_\psi}{dt},\label{bound}
\end{eqnarray}
where $E_{\psi}$ is the average energy of $\ket{\psi}$ relative to the ground state \cite{Hartman:2013qma,Brown:2015bva,Susskind:2014moa,Brown:2015lvg,Pires:2022bmc,deSousa:2025gex,Pires2021,Rangamani:2016dms}. Additionally, the rate of the action depends on whether the black hole is small or large relative to the AdS length scale. These subtleties complicate the understanding of complexity growth in terms of a saturation bound, as proposed in Eq. (\ref{bound}). 

\subsection{First law of thermodynamics}\label{}

The linear growth of complexity during certain dynamical processes can be understood by analyzing two key observations \cite{Hartman:2013qma,Susskind:2014moa}:
\begin{itemize}
   \item First, complexity is an extensive quantity, scaling with the number of active degrees of freedom in the system. A natural measure of the system's size is its thermal entropy, suggesting that both the complexity and its growth rate should be proportional to the system's entropy $\mathcal{S}_{BH}=\mathcal{S}^{\mathcal{Q}}_{BH}+\mathcal{S}^{\partial\mathcal{Q}}_{BH}$ (examples (\ref{ADSBH}), (\ref{BH-BTZ}) and (\ref{BH-EHW})). 

   \item Second, the slope of the linear growth represents a rate which must have dimensions of inverse time or energy \cite{Hartman:2013qma,Susskind:2014moa}. While it might seem plausible to relate the growth rate to the total energy of the system, this approach fails for ground states such as extremal black holes, which possess both entropy and energy but exhibit no time-dependent complexity

\end{itemize}. 

Instead, the correct quantity governing the growth rate is the product of the entropy and temperature of the system (examples (\ref{ADSBH}), (\ref{BH-BTZ}), and (\ref{BH-EHW})):
\begin{eqnarray}
&&{\cal C}_\psi(t)=\frac{2{\cal T}\,\mathcal{S}_{BH}}{\pi\hbar}.\label{TS}
\end{eqnarray}
Within the framework of the AdS/BCFT correspondence, this relationship gains further depth. The selection of times \( t_L \) and \( t_R \) on the left and right asymptotic boundaries of the eternal two-sided Anti-de Sitter (AdS) black hole defines a quantum state \(\ket{\psi(t_L, t_R)}\), expressed as:

\begin{eqnarray}
\ket{\psi(t_L, t_R)} = e^{-i (H_L t_L + H_R t_R)} \ket{\text{TFD}},
\end{eqnarray}
where the Thermofield Double (TFD) state, \(\ket{\text{TFD}} = Z^{-1/2} \sum_\psi e^{-\frac{\beta E_\psi}{2}} \ket{E_\psi}_L \ket{E_\psi}_R\) is a purification of the thermal density matrix on one side of the black hole \cite{Mertens:2022irh,Erdmenger:2022lov,Caceres:2025myu,Erdmenger:2021wzc}. Here, \(Z\) is the partition function, \(\beta\) is the inverse temperature, and \(\ket{E_\psi}_L\), \(\ket{E_\psi}_R\) are the energy eigenstates associated with the left and right boundaries, respectively. Notably, in this convention, time flows upward on both sides of the Penrose diagram Fig. \ref{fig:stream1} \cite{Harlow:2014yka}.

The complexity-action ($\mathcal{C}\mathcal{A}$) duality conjecture posits that the quantum complexity of \(\ket{\psi(t_L, t_R)}\) is proportional to the volume \(\mathcal{V}\) of a maximal spacelike hypersurface anchored at \(t_L\) and \(t_R\) on the AdS boundaries:

\begin{eqnarray}
{\cal C}_\psi(t_L+t_R)=\frac{2{\cal T}\,\mathcal{S}_{BH}}{\pi\hbar},\label{TFDB}
\end{eqnarray}
as the AdS radius for large black holes or the black hole radius ($r_h$) for smaller ones, this framework predicts a linear growth of complexity with time, consistent with expectations from Conformal Field Theory (CFT), and establishes a proportionality to the product of temperature \({\cal T}\) and entropy \(\mathcal{S}_{BH}\) (\ref{TS}) \cite{Brown:2015bva}. As we know, the AdS radius \( L \) is related to the Planck length \( \ell_p \) and the gravitational constant \( G_N \) through the bulk cosmological constant \( \Lambda \) \cite{Brown:2015bva,Susskind:2014moa,Brown:2015lvg}, and the curvature of the AdS spacetime. Thus, we have
\begin{eqnarray}
\mathcal{C}(\ket{\psi(t_L, t_R)})=\frac{2E_{\psi}}{\pi\hbar}\,\,;\,\,E_\psi={\cal T}\,\mathcal{S}_{BH}.\label{TFDB}
\end{eqnarray}
This highlights the relation between quantum mechanics (via \( \hbar \)), gravity (via \( G_N \)), and the geometry of the AdS spacetime (via \( L \)). In the semiclassical limit, where \( \hbar \to 0 \), the radius of AdS \( L \) becomes large, corresponding to a weakly curved bulk spacetime and a strongly coupled boundary CFT \cite{Carmi:2017jqz}. In our prescription, we find that the TFD state is a maximally entangled state of two identical CFTs, which is dual to an eternal black hole in the bulk AdS spacetime. The complexity of the TFD state, as proposed in the ``$complexity=action$'' (\({\cal CA }\)) conjecture, is proportional to the energy of the black hole above the ground state. However, the $\mathcal{C}\mathcal{A}$ duality prescription is not without limitations. The thermodynamic relation $E_\psi={\cal T}\,\mathcal{S}_{BH}$ emerges naturally in our setup, consistent with the first law of black hole thermodynamics \cite{Brown:2015lvg,Sun:2019yps}; the energy \(E_\psi\) of the TFD state is dual to the mass of the black hole.

\begin{figure}[!ht]
\begin{center}
\includegraphics[scale=0.60]{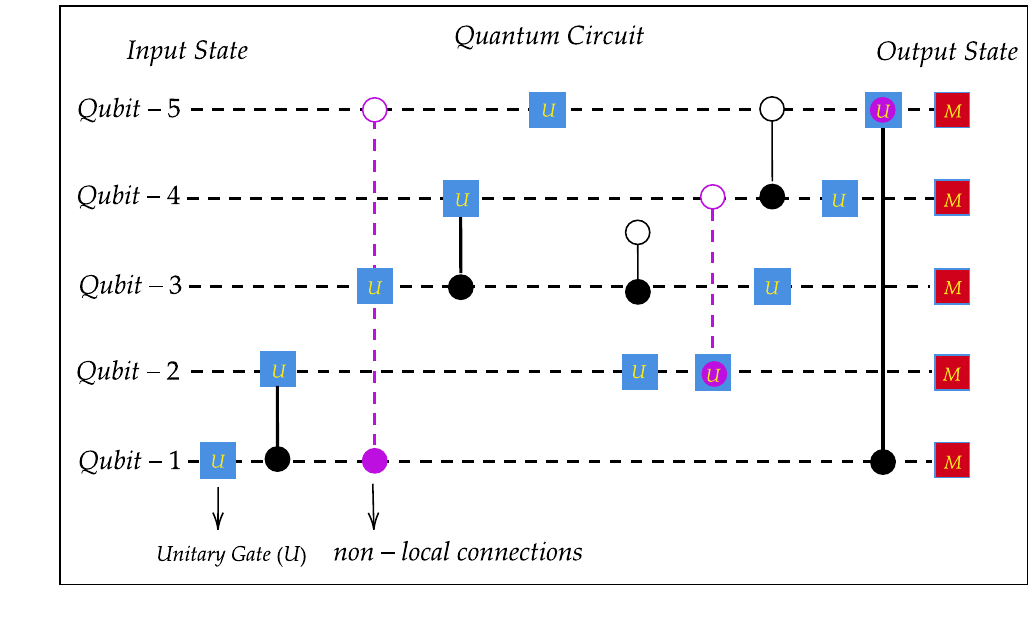}
\caption{The evolution of a quantum circuit implemented through a sequence of quantum gates or a time-dependent Hamiltonian, forming a quantum-in---quantum-out framework.}
\label{QCIR}
\end{center}
\end{figure}

As discussed by \cite{Brown:2015bva,Susskind:2014moa,Brown:2015lvg}, in quantum computation, we define a quantum circuit as a process that begins with a pure input quantum state and evolves to a pure output quantum state Fig. \ref{QCIR}. 

The primary objective of such a computation is to navigate the configuration space of the auxiliary system \(\mathcal{A}\) to reach the desired target state. Importantly, no intermediate measurements are allowed during the computation, as these are not intrinsic to the \(\mathcal{Q}\)-\(\mathcal{A}\) correspondence, where we have an analogy between the complexity of the quantum system \({\cal Q}\) (the quantum system \({\cal Q}\) consists of \({\cal K}\) qubits interacting through a k-local Hamiltonian) and the entropy of the classical system \({\cal A}\) \cite{Brown:2017jil}. Measurements are only performed at the end of the computation to extract useful information. Consequently, computational work and resource requirements are defined solely in terms of the quantum-in---quantum-out process, excluding the final measurement step.

In thermodynamics, the free energy is defined as 

\begin{eqnarray}
{\cal F}={\cal U} - {\cal T}{\cal S}_{BH}. 
\end{eqnarray}
Quantifies the energy available to perform useful work. Extending this concept to the auxiliary system, we define the free energy as
\begin{eqnarray}
{\cal F}_{{\cal A}} ={\cal U}_{{\cal A}}-{\cal T}_{{\cal A}}{\cal S}_{BH}\,,
\end{eqnarray}
with \({\cal T}_{{\cal A}}{\cal S}_{BH}=\pi\hbar{\cal C}_\psi(t)/2\), where \({\cal C}_\psi(t)\) represents the complexity of the system; \({\cal U}_{{\cal A}}\) are fixed parameters that depend on the number of qubits, as determined by the relation \({\cal U}_{{\cal A}}={\cal K}/2\). The complexity \({\cal C}_\psi(t)\) emerges as the only variable in the free energy, suggesting that \(-{\cal C}_\psi(t)\) can be interpreted as a resource analogous to the free energy in thermodynamics \cite{Sun:2019yps,Brown:2017jil}. The gates in the quantum circuit Fig. \ref{QCIR} remain unitary regardless of the thermodynamic relation \(\mathcal{C}_\psi(t)\). Nevertheless, the number of gates required to satisfy this relation depends on the values of \(\mathcal{T}_{{\cal A}}\) and \(\mathcal{S}_{BH}\). In particular, when either \(\mathcal{T}_{{\cal A}}\) or \(\mathcal{S}_{BH}\) increases, a greater number of gates is necessary to achieve the desired complexity growth rate, whereas lower values of these quantities reduce the number of gates needed. To account for cases where \(\mathcal{C}_\psi(t)\) increases, additional unitary gates have been incorporated, reflecting the growth in complexity. The purple dotted lines in Fig. \ref{QCIR} represent non-local connections induced by \(\mathcal{C}_\psi(t)\), linking distant qubits and thereby introducing non-local effects. Furthermore, extra controlled operations have been implemented to capture the intricate correlations within the system. This behavior is consistent with the holographic principle, according to which the complexity of the boundary theory mirrors the dynamics of the bulk spacetime \cite{Sun:2019yps,Brown:2017jil,Fan:2019mbp}. Specifically, the free energy \({\cal F}_\psi\) can be linked to the holographic complexity proposals, where \({\cal C}_\psi(t)\) is computed using the conjectures ``$complexity = volume$'' or ``$complexity = action$". These conjectures provide a geometric interpretation of complexity in terms of bulk quantities, such as the volume of a maximal slice or the on-shell action evaluated within the AdS spacetime.

\begin{figure}[!ht]
\begin{center}
\includegraphics[scale=0.80]{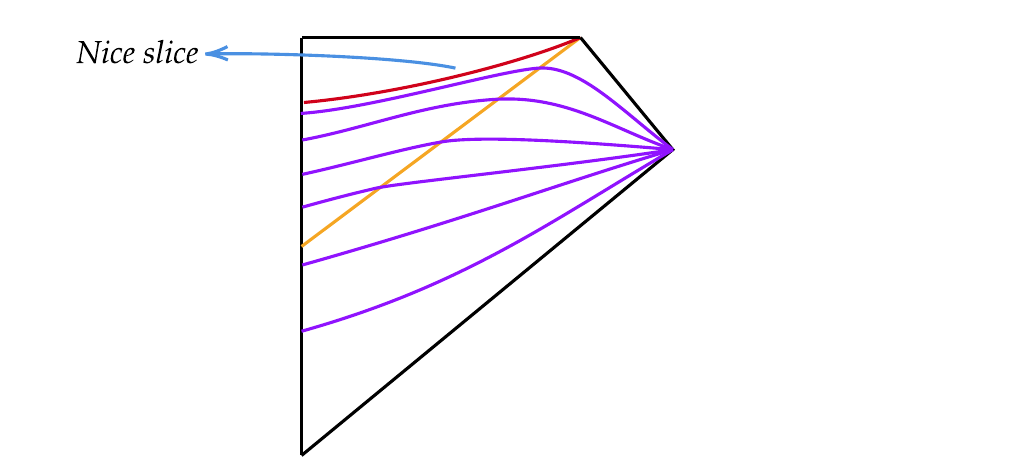}
\caption{The Penrose diagram is depicted with a foliation of spacetime by maximal volume slices, showing the evolution of the geometry. The red curve represents the final maximal slice, which is the anchor for the computation of holographic complexity \cite{Brown:2015bva,Harlow:2014yka}.}
\label{CPX}
\end{center}
\end{figure}

The choice of maximal-volume slices is inherently nonunique, and these slices do not fully foliate the region behind the event horizon \cite{Hartman:2013qma,Brown:2015bva,Susskind:2014moa,Brown:2015lvg,Harlow:2014yka}. On the other hand, a nice slice is the red curve in Fig. \ref{CPX}, where this approach aligns with the complexity duality framework, where maximal slices play a central role in connecting the bulk geometry to boundary quantum states \cite{Brown:2015bva,Harlow:2014yka}. 

Additionally, the introduction of the length scale \(L\) introduces an element of arbitrariness, as its definition varies depending on the specific black hole geometry under consideration. Our refined framework addresses these ambiguities while preserving the core advantages of the $\mathcal{C}\mathcal{A}$ duality hypothesis, offering a more robust and universal approach to the study of holographic complexity \cite{Agon:2018zso}.

\subsection{Testing shock waves}\label{TSHO}

The complexity/geometry duality has undergone extensive testing, particularly in scenarios involving shock waves and their impact on the growth of complexity and spacetime volume \cite{Brown:2015lvg,Roberts:2014ifa,Shenker:2013pqa,Susskind:2014rva}. To extend this framework, we analyze the growth of complexity in eternal black hole geometries perturbed by shock waves, now within the context of the AdS/BCFT correspondence and Horndeski gravity. The inclusion of Horndeski terms, parameterized by $\alpha$ and $\gamma$, introduces novel modifications to the bulk geometry and boundary dynamics, offering a richer structure for studying complexity.

Shock waves are generated by perturbing the thermofield double (TFD) state with thermal-scale operators \cite{Roberts:2014isa,Caceres:2023gfa}, expressed as:
\begin{eqnarray}
e^{-i H_L t_L} e^{-i H_R t_R} W(t_n) \dots W(t_1) \ket{\text{TFD}},
\end{eqnarray}
where \( W(t) = e^{i H_L t} W e^{-i H_L t} \) represents a smeared operator acting on the left boundary. These states correspond to eternal black hole geometries perturbed by multiple shock waves, with the perturbations propagating along null directions in the bulk.  

\begin{figure}[!ht]
\begin{center}
\includegraphics[scale=0.85]{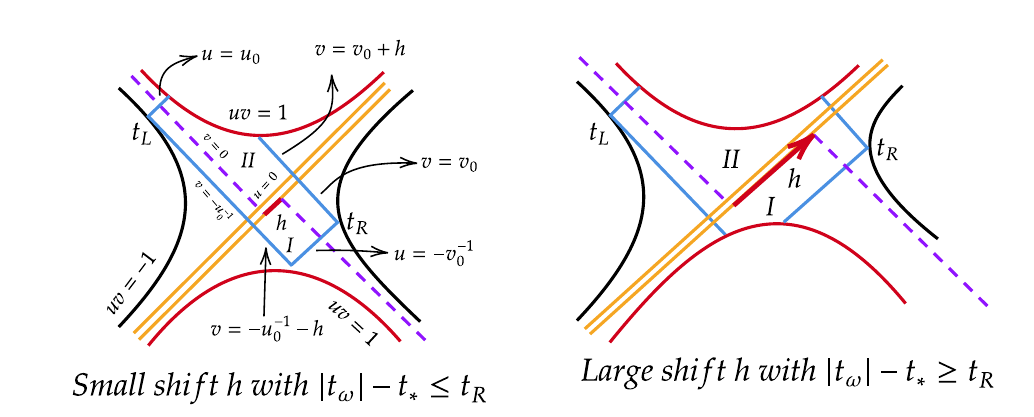}
\caption{Kruskal diagram: The blue lines delineate the boundary of the wedge region $\mathcal{W}$, while the orange lines shaded area, formed by the intersection of $\mathcal{W}$ and the black hole interior, encapsulates the region where the gravitational action contributes to the holographic complexity. The regions along $u=0$ are the shock wave.}
\label{SHOC}
\end{center}
\end{figure}

The growth of complexity in these geometries provides a stringent test of the complexity equals action (${\cal\,CA}$) conjecture. By introducing shock waves at times \( t_1, \dots, t_n \), we can explore the interplay between boundary time evolution and bulk dynamics. In Horndeski gravity, the parameters $\alpha$ and $\gamma$ modify the causal structure of the Wheeler-DeWitt (WdW) patch, and the rate of complexity growth. For instance, we consider the black hole in Kruskal coordinates:

\begin{eqnarray}
dS^2_{\mathcal{Q}}=-\frac{36f(r)h(x)\delta\,(u)du^2}{r^2_huv}-\frac{36f(r)dudv}{r^2_h u v}+r^2 (dx^2+dy^2),\label{Kruscalm}
\end{eqnarray}
In Horndeski gravity, perturbations do not generally propagate on the null cone of the background metric, and the characteristic hypersurfaces can be spacelike \cite{Kovacs:2020pns,Papallo:2017qvl,Bettoni:2016mij}. For the class of backgrounds and couplings considered here, we assume that the dominant modes associated with the shock wave can be effectively described by null hypersurfaces of the Kruskal metric (\ref{Kruscalm}). In other words, we treat the shock as propagating along \(u = 0\) with an effective speed that matches the null structure of this metric. We know that the relationship between Kruskal coordinates and Schwarzschild coordinates is given by
\begin{eqnarray}
uv = - e^{\frac{4\pi}{\beta}r_*(r)}, \qquad u/v = - e^{-\frac{4\pi}{\beta} t},
\end{eqnarray}
with $dr_* = dr / f(r)$. The expectation value of the stress tensor \cite{Roberts:2014ifa,Roberts:2014isa,Caceres:2023gfa} in the state \(u=0\) is 
\begin{eqnarray}
T_{uu}=\frac{3(\alpha+\gamma\Lambda)}{2L^2\kappa}\Omega\,e^{\frac{2\pi}{\beta}|t_{\omega}|}\delta(u)
\end{eqnarray}
\(\Omega\) is the dimensionless asymptotic energy of the perturbation \cite{Roberts:2014isa}. Using the Horndeski equation of motion ${\cal E}^{\mathcal{Q}}_{\mu\nu}=0$, we find 
\begin{eqnarray}
(-\partial_{i}\partial_{i}+2\Lambda)h(x)=\frac{3(\alpha+\gamma\Lambda)}{2L^2\kappa}\Omega\,e^{\frac{2\pi}{\beta}|t_{\omega}|}h(x),
\end{eqnarray}
assuming a thermal-scale initial energy $\Omega$ and expanding for large $|x|$, we have the solution
\begin{eqnarray}
h(x)=e^{\frac{2\pi}{\beta}(|t_w|-t_* )-2\Lambda|x|} \label{eq:shift}
\end{eqnarray}
where we have defined the fast scrambling time $t_* = \frac{\beta}{2\pi} \log \frac{2\kappa\,L^2} {3(\alpha+\gamma\Lambda)}$. 
The Wheeler-DeWitt patch, which determines the action and hence the complexity, undergoes a transition depending on the relative magnitudes of \( |t_w| - t_* \) and \( t_R \). For small shifts (\( |t_w| - t_* \leq t_R \)) Fig. \ref{SHOC}, the complexity grows linearly with boundary time, as the backreaction of the shock wave is negligible. However, for large changes (\( |t_w| - t_* \geq t_R \)), the backreaction becomes significant, and the complexity growth rate doubles due to the contribution of the shock wave. This behavior is consistent with the "switchback effect" \cite{Susskind:2014rva}, where the delay in complexity growth is attributed to the scrambling dynamics of the perturbation \cite{Roberts:2014ifa,Roberts:2014isa}.

The shock wave, represented by the double orange lines along $u=0$ Fig. \ref{SHOC}, introduces a discontinuity that propagates through the spacetime, changing the causal structure and the complexity of the system. The geometric structure of the region under consideration, particularly whether it intersects the past singularity, is governed by the boundary times $t_L$, $t_R$, and the shift parameter $h$ Fig. \ref{SHOC}. Specifically, for $u_0^{-1} + h < v_0$, the past-directed light rays converge within the bulk interior, while for $u_0^{-1} + h > v_0$, they terminate at the singularity. To formalize this, we employ the Kruskal coordinate definitions:

\[
u_0 = e^{\frac{2\pi}{\beta} t_L}, \qquad v_0 = e^{\frac{2\pi}{\beta} t_R}.
\]

Here, $t_L > 0$ ensures that $u_0^{-1}$ is exponentially suppressed, allowing us to neglect it in subsequent calculations. This simplification leads to the condition for the intersection of the causal wedge $\mathcal{W}$ with the past singularity, expressed as $|t_w| - t_* \geq t_R$. Now, we compute the action for the region $II$ with arbitrary $\mathcal{W}$ edges $u_1$ and $v_1$, under the assumption that $t_L, t_R, |t_w| > 0$. This allows us to safely take the limit $u_1, v_1 \gg 1$. The bulk contribution to the action is straightforward and yields:

\begin{eqnarray}
{\cal A}_{bulk}=\frac{2r^2_h\Delta_{{\cal\,Q}}}{\kappa}\left(1+\frac{\xi}{4}\right)\log(v_1\,u_1).
\end{eqnarray}
The boundary contributions arise from the horizon and the singularity, while the edges of $\mathcal{W}$ do not contribute. The boundary action is given by:
\begin{eqnarray}
{\cal A}_{bdry}=\frac{2r^2_h\Delta_{{\cal\partial\,Q}}}{\kappa}\left(1+\frac{\xi}{4}\right)\log(v_1\,u_1).
\end{eqnarray}
The total action (${\cal A}_{II}={\cal A}_{bulk}+{\cal A}_{bdry}$) is for the region $II$. To analyze the one-shock geometry with a small shift $|t_w| - t_* \leq t_R$ (as illustrated on the left side of Fig. \ref{SHOC}, we focus on the top five-sided region $II$. Replacing $u_1 = u_0$ and $v_1 = v_0 + h$ into ${\cal A}_{II}$, and using the condition $v_0 > h$, we find that the dominant contribution to the action arises from the logarithmic dependence on $u_1 v_1$. This dependence is further modified by the parameters $\alpha$ and $\gamma$, which encode the effects of Horndeski gravity on the shockwave geometry and in the holographic dual. With this result in hand, we now analyze the one-shock geometry in the regime of small shifts, characterized by $|t_w| - t_* \leq t_R$ (as depicted on the left side of Fig. \ref{SHOC}). In this scenario, the dominant contribution comes from the top region of the side, denoted region $II$. Using $u_1 = u_0$ and $v_1 = v_0 + h$ into the action integral for region $II$, as given in ${\cal A}_{II}$, and noting that $v_0 > h$, we obtain the complexity contribution:
\begin{eqnarray}
&&{\cal A}_{II}=\left[\frac{2r^2_h\Delta_{{\cal\,Q}}}{k^2}\left(1+\frac{\xi}{4}\right)+\frac{2r^2_h\Delta_{{\cal\partial\,Q}}}{k^2}\left(1+\frac{\xi}{4}\right)\right]\frac{2\pi}{\beta}(t_R+t_L),\\
&&{\cal A}_{II}=(\mathcal{S}^{\mathcal{Q}}_{BH}+\mathcal{S}^{\partial\mathcal{Q}}_{BH})\frac{2\pi}{\beta}(t_R+t_L).
\end{eqnarray}
In terms of temperature, we have the following:
\begin{eqnarray}
{\cal A}_{II}={\cal T}\mathcal{S}_{BH}(t_R+t_L)\to\mathcal{C}_{II}(t_R+t_L)=\frac{2{\cal T}\,\mathcal{S}_{BH}}{\pi\hbar},
\end{eqnarray}
which is identical to the result (\ref{ADSBH}), (\ref{BH-BTZ}), and (\ref{BH-EHW}) in the absence of the shock wave. This behavior can be understood as a consequence of the negligible backreaction of the perturbation in the large $t_R$ limit, where the system evolves into a regime dominated by the thermofield double state \cite{Brown:2015lvg,Roberts:2014ifa,Shenker:2013pqa,Susskind:2014rva}. 

\section{Conclusions and discussions}\label{CONC}

This work provides an investigation of the possible connection between quantum complexity and gravitational dynamics within the framework of Horndeski gravity, extending the AdS/BCFT correspondence to include scalar-tensor interactions. The detailed derivation of black hole entropy in Horndeski gravity employs three complementary approaches: the Wald entropy formula, the Wald formalism, and the holographic renormalization scheme. These methods reveal that the entropy includes corrections proportional to the squared norm of the scalar field's gradient, emphasizing the role of scalar-tensor couplings in black hole thermodynamics \cite{Caceres:2017lbr,Minamitsuji:2013vra,Callan:2012ip}. Our analysis shows the consistency with the first law of thermodynamics and provides a robust framework for studying the thermodynamic properties of black holes in modified gravity scenarios. We also examine the relation between entanglement entropy, phase transitions, and the growth of complexity. In our numerical applications to 3D planar and spherical black holes, we reveal the connection between bulk geometry and boundary degrees of freedom. The phase transition between connected and disconnected minimal surfaces provides a geometric manifestation of the transition from linear complexity growth to saturation. This transition is particularly significant in the context of Horndeski gravity, where the coupling parameter $\gamma$ introduces modifications to the functional entanglement entropy, the boundary entropy, and the critical angle of the phase transitions.

In our refining of the ``complexity = action" (${\cal\,CA}$) conjecture \cite{Susskind:2014rva,Nielsen:2005mkt}, we show the robustness and universality of this conjecture across black hole configurations, including rotating, charged, and planar systems. The inclusion of angular momentum and electric charge introduces additional corrections to the Wheeler-DeWitt (WdW) patch, impacting the rate of complexity growth and its holographic interpretation. For rotating black holes, we confirm that the ${\cal\,CA}$ conjecture remains valid, with the rate of complexity growth proportional to the product of black hole entropy and temperature. Similarly, the extension of the ${\cal\,CA}$ conjecture to the Einstein-Horndeski-Maxwell action incorporates gauge fields, providing a richer framework for exploring the dynamics of charged black holes. The results establish a direct connection between holographic complexity and the thermodynamic properties of black holes, particularly in the presence of an electric field.

Finally, we have tested the impact of shock waves on complexity growth \cite{complexityshocks,Susskind:2014moa}, confirming the persistence of the ``switchback effect" in the presence of Horndeski modifications. This backreaction of shock waves introduces corrections to the causal structure of the WdW patch, offering new insights into the interplay between boundary time evolution and bulk dynamics. These findings underscore the robustness of the ${\cal\,CA}$ conjecture in capturing the holographic complexity of black holes, even in the presence of perturbations \cite{complexityshocks,Susskind:2014moa,Susskind:2014rva,Brown:2015bva,Brown:2015lvg}. The results underscore the universality of the ${\cal\,CA}$ conjecture and its applicability to a wide range of gravitational theories, paving the way for future research into the interplay between quantum information and gravitational dynamics.

\acknowledgments
The authors would like to thank Hao Geng and Diego Paiva Pires for the fruitful discussions. ENS acknowledges the contribution of the LISA CosWG, and of COST Actions CA21136 ``Addressing observational tensions in cosmology with systematics and fundamental physics (CosmoVerse)'' and CA23130 ``Bridging high
and low energies in search of quantum gravity (BridgeQG)''. Fabiano F. Santos is partially supported by Conselho Nacional de Desenvolvimento Cient\'{\i}fico e Tecnol\'{o}gico (CNPq) under grant 302835/2024-5.

\end{document}